\documentclass{emulateapj}
\usepackage{epsfig}
\usepackage[T1]{fontenc}
\usepackage{epstopdf}
\usepackage{natbib}
\usepackage{graphicx}
\usepackage{xspace}
\usepackage{hyperref}
\newcommand{\specialcell}[2][c]{%
  \begin{tabular}[#1]{@{}c@{}}#2\end{tabular}}

\newcommand{\nc}[2]{\newcommand{#1}{\ensuremath{#2}\xspace}}
\newcommand{\num}[2]{\newcommand{#1}{{#2}\xspace}}

\nc{\Kepler}{ \textit{Kepler} }
\nc{\Kp}{ \textit{Kp} }
\nc{\SpecMatch}{\mathrm{SpecMatch}} 
\nc{\SME}{\mathrm{SME@XSEDE}} 

\nc{\Rsun}{ R_{\odot} }
\nc{\Rearth}{ R_{\oplus} }
\nc{\Mearth}{ M_{\oplus} }
\nc{\Rp}{ R_P }
\nc{\Mp}{ M_P }
\nc{\Rstar}{R_\star} 
\nc{\Mstar}{M_\star}
\newcommand{\caii}{\ion{Ca}{2}~H \& K}

\nc{\teff}{\ensuremath{T_{\rm eff}}}
\nc{\logg}{\ensuremath{\log{g}}}
\nc{\feh}{\ensuremath{\mathrm{[Fe/H]}}}
\nc{\vsini}{\ensuremath{V \sin i}}

\nc{\kms}{\text{km\,s}^{-1}}
\nc{\ms}{\text{m\,s}^{-1}}
\nc{\K}{\ensuremath{\mathrm{K}}}
\nc{\dex}{\ensuremath{\mathrm{dex}}}

\nc{\nstars}{1305} 
\nc{\nstarsMaglim}{960} 
\nc{\nstarsMulti}{484} 
\nc{\nstarsHZ}{127} 
\nc{\nstarsUSP}{71} 
\nc{\nstarsFP}{113} 
\nc{\nstarsOther}{38} 
\nc{\nstarsSX}{972} 
\nc{\nstarsDisagree}{26} 

\nc{\cksSigTeff}{60} 
\nc{\cksSiglogg}{0.10} 
\nc{\cksSigfeh}{0.04} 

\nc{\nplanets}{2075} 
\nc{\nplanetsMaglim}{1385} 
\nc{\nplanetsMulti}{1254} 
\nc{\nplanetsHZ}{127} 
\nc{\nplanetsUSP}{71} 
\nc{\nplanetsFP}{175} 
\nc{\nplanetsOther}{38} 

\nc{\BrewerNstarscommon}{106} 
\nc{\BrewerTeffOffset}{0} 
\nc{\BrewerTeffRMS}{61} 
\nc{\BrewerLoggOffset}{0.000} 
\nc{\BrewerLoggRMS}{0.099} 
\nc{\BrewerFehOffset}{0.00} 
\nc{\BrewerFehRMS}{0.06} 

\nc{\SXNstarscommon}{934} 
\nc{\SXTeffOffset}{1} 
\nc{\SXTeffRMS}{68} 
\nc{\SXLoggOffset}{0.00} 
\nc{\SXLoggRMS}{0.09} 
\nc{\SXFehOffset}{0.001} 
\nc{\SXFehRMS}{0.036} 

\nc{\HuberBNstarscommon}{71} 
\nc{\HuberBNplanetscommon}{xxx} 
\nc{\HuberBTeffOffset}{-10} 
\nc{\HuberBTeffRMS}{59} 
\nc{\HuberBLoggOffset}{-0.03} 
\nc{\HuberBLoggRMS}{0.08} 
\nc{\HuberBFehOffset}{-0.04} 
\nc{\HuberBFehRMS}{0.09} 

\nc{\BrunttSMNstarscommon}{57} 
\nc{\BrunttSMNplanetscommon}{xxx} 
\nc{\BrunttSMTeffOffset}{-3} 
\nc{\BrunttSMTeffRMS}{70} 
\nc{\BrunttSMLoggOffset}{\phantom{-}0.02} 
\nc{\BrunttSMLoggRMS}{0.11} 
\nc{\BrunttSMFehOffset}{-0.056} 
\nc{\BrunttSMFehRMS}{0.056} 

\nc{\KICNstarscommon}{1215} 
\nc{\KICNplanetscommon}{xxx} 
\nc{\KICTeffOffset}{-52} 
\nc{\KICTeffRMS}{161} 
\nc{\KICLoggOffset}{+0.09} 
\nc{\KICLoggRMS}{0.29} 
\nc{\KICFehOffset}{-0.194} 
\nc{\KICFehRMS}{0.254} 

\nc{\HuberNstarscommon}{1302} 
\nc{\HuberNplanetscommon}{xxx} 
\nc{\HuberTeffOffset}{+128} 
\nc{\HuberTeffRMS}{193} 
\nc{\HuberLoggOffset}{+0.03} 
\nc{\HuberLoggRMS}{0.26} 
\nc{\HuberFehOffset}{-0.148} 
\nc{\HuberFehRMS}{0.227} 

\nc{\LAMOSTNstarscommon}{283} 
\nc{\LAMOSTNplanetscommon}{xxx} 
\nc{\LAMOSTTeffOffset}{-7} 
\nc{\LAMOSTTeffRMS}{113} 
\nc{\LAMOSTLoggOffset}{-0.03} 
\nc{\LAMOSTLoggRMS}{0.14} 
\nc{\LAMOSTFehOffset}{-0.053} 
\nc{\LAMOSTFehRMS}{0.119} 

\nc{\BuchhaveNstarscommon}{396} 
\nc{\BuchhaveNplanetscommon}{xxx} 
\nc{\BuchhaveTeffOffset}{-5} 
\nc{\BuchhaveTeffRMS}{93} 
\nc{\BuchhaveLoggOffset}{+0.02} 
\nc{\BuchhaveLoggRMS}{0.15} 
\nc{\BuchhaveFehOffset}{+0.031} 
\nc{\BuchhaveFehRMS}{0.117} 

\nc{\EndlNstarscommon}{44} 
\nc{\EndlNplanetscommon}{xxx} 
\nc{\EndlTeffOffset}{+79} 
\nc{\EndlTeffRMS}{70} 
\nc{\EndlLoggOffset}{+0.05} 
\nc{\EndlLoggRMS}{0.15} 
\nc{\EndlFehOffset}{-0.053} 
\nc{\EndlFehRMS}{0.106} 

\nc{\EverettNstarscommon}{143} 
\nc{\EverettNplanetscommon}{xxx} 
\nc{\EverettTeffOffset}{-40} 
\nc{\EverettTeffRMS}{102} 
\nc{\EverettLoggOffset}{\phantom{-}0.05} 
\nc{\EverettLoggRMS}{0.18} 
\nc{\EverettFehOffset}{+0.008} 
\nc{\EverettFehRMS}{0.076} 

\nc{\BastienNstarscommon}{232} 
\nc{\BastienNplanetscommon}{xxx} 
\nc{\BastienLoggOffset}{-0.11} 
\nc{\BastienLoggRMS}{0.21} 

\num{\TeffDiffMed}{37} 
\num{\TeffDiffMedM}{24.9} 
\num{\TeffDiffMedP}{92.3} 
\num{\LoggDiffMed}{$-$0.0037} 
\num{\feCoeffZero}{0.0543} 
\num{\feCoeffOne}{0.101}   

\shortauthors{Petigura {et~al.}}
\shorttitle{CKS I. High-Resolution Spectroscopy of 1305 Stars Hosting {\it Kepler} Transiting Planets }
\submitted{Submitted to AAS Journals}
\begin{document}
\pagenumbering{arabic}


\title{The California-Kepler Survey. \\
       I. High Resolution Spectroscopy of 1305 Stars Hosting \textit{Kepler} Transiting Planets\altaffilmark{1}}
\author{
Erik A.\ Petigura\altaffilmark{2,11}, 
Andrew W.\ Howard\altaffilmark{2,3}, 
Geoffrey W.\ Marcy\altaffilmark{4},
John Asher Johnson\altaffilmark{5},
Howard Isaacson\altaffilmark{4}, 
Phillip A.\ Cargile\altaffilmark{5},
Leslie Hebb\altaffilmark{6}, 
Benjamin J.\ Fulton\altaffilmark{3,2,12}, 
Lauren M.\ Weiss\altaffilmark{7,13}, 
Timothy D.\ Morton\altaffilmark{9}, 
Joshua N.\ Winn\altaffilmark{9}, 
Leslie A.\ Rogers\altaffilmark{10}, 
Evan Sinukoff\altaffilmark{3,2,14}, 
Lea A.\ Hirsch\altaffilmark{4}, 
Ian J.\ M.\ Crossfield\altaffilmark{8,15}
}
\altaffiltext{1}{Based on observations obtained at the W.\,M.\,Keck Observatory, 
                 which is operated jointly by the University of California and the 
                 California Institute of Technology.  Keck time was granted for this project by 
                 the University of California, and California Institute of Technology, the University of Hawaii, and NASA.} 
\altaffiltext{2}{California Institute of Technology, Pasadena, CA, 91125, USA}
\altaffiltext{3}{Institute for Astronomy, University of Hawai`i at M\={a}noa, Honolulu, HI 96822, USA}
\altaffiltext{4}{Department of Astronomy, University of California, Berkeley, CA 94720, USA}
\altaffiltext{5}{Harvard-Smithsonian Center for Astrophysics, 60 Garden St, Cambridge, MA 02138, USA}
\altaffiltext{6}{Hobart and William Smith Colleges, Geneva, NY 14456, USA}
\altaffiltext{7}{Institut de Recherche sur les Exoplan\`{e}tes, Universit\'{e} de Montr\'{e}al, Montr\'{e}al, QC, Canada}
\altaffiltext{8}{Astronomy and Astrophysics Department, University of California, Santa Cruz, CA, USA}
\altaffiltext{9}{Department of Astrophysical Sciences, Peyton Hall, 4 Ivy Lane, Princeton, NJ 08540, USA}
\altaffiltext{10}{Department of Astronomy \& Astrophysics, University of Chicago, 5640 South Ellis Avenue, Chicago, IL 60637, USA}

\altaffiltext{11}{Hubble Fellow}
\altaffiltext{12}{National Science Foundation Graduate Research Fellow}
\altaffiltext{13}{Trottier Fellow}
\altaffiltext{14}{Natural Sciences and Engineering Research Council of Canada Graduate Student Fellow}
\altaffiltext{15}{NASA Sagan Fellow}

\begin{abstract}
The California-Kepler Survey (CKS) is an observational program to improve our knowledge of the properties of stars found to host transiting planets by NASA's \Kepler Mission.  The improvement stems from new high-resolution optical spectra obtained using HIRES at the W.\ M.\ Keck Observatory. The CKS stellar sample comprises \nstars stars classified as {\it Kepler} Objects of Interest, hosting a total of \nplanets transiting planets.  The primary sample is magnitude-limited ($Kp < 14.2$) and contains \nstarsMaglim stars with \nplanetsMaglim planets.  The sample was extended to include some fainter stars that host multiple planets, ultra short period planets, or habitable zone planets. The spectroscopic parameters were determined with two different codes, one based on template matching and the other on direct spectral synthesis using radiative transfer. We demonstrate a precision of \cksSigTeff~K in \teff, \cksSiglogg~dex in \logg, \cksSigfeh~dex in \feh, and 1.0 \kms in \vsini. In this paper, we describe the CKS project and present a uniform catalog of spectroscopic parameters. Subsequent papers in this series present catalogs of derived stellar properties such as mass, radius and age; revised planet properties; and statistical explorations of the ensemble. CKS is the largest survey to determine the properties of \Kepler stars using a uniform set of high-resolution, high signal-to-noise ratio spectra. The HIRES spectra are available to the community for independent analyses.  
\end{abstract}


\section{Introduction}
\label{sec:intro}

The NASA \Kepler Mission \citep{Borucki2010,Koch2010,Borucki2016} has ushered in a new era in astronomy, in which extrasolar planets are known to be ubiquitous. The canon of \Kepler papers contains descriptions of many remarkable planetary systems. The precision of \Kepler photometry enabled the detection of planets as small as Mercury \citep{Barclay2013a}, and the long, nearly uninterrupted dataset revealed a plethora of compact systems of multiple transiting planets (e.g. Kepler-11; \citealt{Lissauer2011}). These iconic \Kepler systems present opportunities to determine planet masses and orbital properties through dynamical effects \citep{Ford2011,Lissauer2011c} and have inspired new classes of planet formation models \citep{Hansen2012,Chiang2013}.  Circumbinary planets were found \citep{Doyle2011}, and searches for moons \citep{Kipping2012} and rings \citep{Heising2015} were attempted. \Kepler also revealed planets resembling the Earth in size and incident stellar flux \citep{Borucki2012,Borucki2013,Quintana2014,Torres2015}.

Doppler measurements of the masses of Kepler-discovered planets provided constraints on the composition of small planets extending down to the size of Earth \citep[e.g., Kepler-78b;][]{Howard2013_kepler78,Pepe2013}. Once the sample of such measurements was large enough, patterns began to emerge. \cite{Marcy2014} measured the masses of 49 planets and found evidence for a transition from rock- to gas-dominated compositions with increasing planet size \citep{Weiss2014, Rogers2015, Wolfgang2015}.

The \Kepler canon also includes statistical analyses of the properties of thousands of transiting planets and their host stars. Shortly before the launch of \Kepler, radial-velocity (RV) surveys found that the occurrence of close-in (< 0.5~AU) planets around FGK stars rises rapidly with decreasing mass, with Neptune-mass planets outnumbering Jovian mass planets \citep{Howard2010, Mayor2011}. After just a few months of \Kepler photometry, the prevalence of planets smaller than Neptune (\Rp $<$ 4.0 \Rearth) was confirmed and came into sharper focus. Many studies quantified the occurrence of planets as a function of planet radius and orbital period \citep{Howard2012_kepocc, Petigura2013_apj, Fressin2013,Dressing2013}. Further work showed that Earth-size planets are common in and near the habitable zone \citep{Petigura2013_pnas, Dressing2015, Burke_2015}.

A important limiting factor in large statistical analyses of \Kepler planets is the quality of the host star properties. Using only broadband photometry, the Kepler Input Catalog \citep[KIC;][]{Brown2011_kic} provided stellar effective temperatures and radii good to about 200 K and 30\%. These parameters limit precision planet size and incident stellar flux measurements, obscuring important features. For example, any fine details in the radius distribution of planets are smeared out by the uncertainties associated with photometric stellar radii.

This paper introduces the California-Kepler Survey (CKS), a large observational campaign to measure the properties of \Kepler planets and their host stars. CKS is designed in the same spirit as the pioneering spectroscopic surveys of nearby stars targeted in Doppler planet searches \citep{Valenti05}.  By providing a large sample of well-characterized stars, those early surveys mapped out the strong correlation between giant-planet occurrence and stellar metallicity \citep{Fischer2005} and planet occurrence as a function of planet mass, stellar mass, and orbital distance \citep{Cumming2008,Howard2010,Johnson2010}.  

For the CKS project we measure stellar parameters and conduct statistical analyses of the \Kepler planet population.  
A central motivation for CKS was to reduce the uncertainty in the sizes of \Kepler stars and planets 
from typically 30\% in the KIC to 10\% using high-resolution spectroscopy.
With this improvement, CKS enables more powerful and discriminating statistical studies of the occurrence of planets as a function of the properties of the planet and the host star, including its mass, age, and metallicity.

The CKS project grew out of experience with the \Kepler Follow-up Observation Program (KFOP; \citealt{Gautier2010}), which carried out extensive ground-based observations of hundreds of \Kepler Objects of Interest (KOIs) using many facilities operated by dozens of astronomers.\footnote{This effort was later enlarged to include any willing observers and renamed the Community Follow-up Observing Program (CFOP).} These observations included direct imaging \citep{Adams2012,Adams2013,Baranec2016,Ziegler2017,Furlan2017} as well as high-resolution spectroscopy \citep{Gautier2012,Everett2013,Buchhave2012,Buchhave_2014}.
The \textit{Spitzer Space Telescope} was also used for characterization of \Kepler-discovered planets \citep{Desert2015}.

\begin{deluxetable*}{l}[tb]
\tabletypesize{\footnotesize}
\tablecaption{Papers from the California Kepler Survey
\label{tab:papers}}
\tablewidth{0pt}
\tablehead{
}
\startdata
\sidehead{Primary CKS Papers}
~~~~CKS I.\ High-Resolution Spectroscopy of 1305 Stars Hosting {\it Kepler} Transiting Planets (this paper) \\ 
~~~~CKS II.\ Precise Physical Properties of 2075 {\it Kepler} Planets and Their Host Stars (Johnson et al., submitted)\\
~~~~CKS III.\ A Gap in the Radius Distribution of Small Planets (Fulton et al., submitted)\\
~~~~CKS IV.\ Metallicities of \Kepler Planet Hosts (Petigura et al., to be submitted)\\
~~~~CKS V.\ Stellar and Planetary Properties of Kepler Multiplanet Systems (Weiss et al., to be submitted)\\
\sidehead{Related Papers Using CKS Data}
~~~~Detection of Stars Within $\sim$0.8\arcsec of Kepler Objects of Interest \citep{Kolbl2015}\\
~~~~Absence of a Metallicity Effect for Ultra-short-period Planets (Winn et al.\ 2017, submitted)\\
~~~~Identifying Young Kepler Planet Host Stars from Keck-HIRES Spectra of Lithium (Berger et al., in prep)
\enddata
\end{deluxetable*}

In this paper, we describe the survey (Sec.\ \ref{sec:cks}), the spectroscopic pipelines (Sec.\ \ref{sec:pipelines}), the catalog of spectroscopic parameters (Sec.\ \ref{sec:catalog}), a comparison of results from other surveys (Sec.\ \ref{sec:comparison}), and a summary of conclusions (Sec.\ \ref{sec:summary}).  Table \ref{tab:papers} outlines the papers in the CKS series.
  Paper II presents the  stellar radii, masses, and approximate ages for stars in the CKS sample, based on the spectroscopic parameters presented here.  Papers III, IV, and V are statistical analyses of planet and star properties enabled by this large and precise catalog.  A set of related papers make use of the CKS data to conduct complementary analyses.

\section{The California-Kepler Survey}
\label{sec:cks}

\subsection{Project Plan}
\label{sec:plan}

The original goal of the CKS project was to measure the stellar properties of all 997 host stars in the first large \Kepler planet catalog \citep{Borucki2011}.  As the \Kepler planet catalogs grew in size \citep{Batalha2013,Burke2014}, we decided on a magnitude limit of $Kp< 14.2$ (\Kepler apparent magnitude) for the primary CKS sample. Most of the spectra were collected during the 2012, 2013, and 2014 observing seasons. During this time the tabulated `dispositions' of some KOIs changed between `candidate', `confirmed', `validated', and `false positive'. We discuss the dispositions that we adopted in Sec.\ \ref{sec:false_positives}.  Planet candidates have low probabilities of being false positives, typically $<$10\% \citep{Morton2011}.  For simplicity, we refer to KOIs as ``planets'' throughout much of this paper, except when describing known false positives.

The CKS project is independent from the KFOP observations that were in direct support of the \Kepler mission. CKS observations of the magnitude-limited sample (see Sec.\ \ref{sec:stellar_samples}) were conducted using Keck time granted for this project by the University of California, the California Institute of Technology, and the University of Hawaii.  Observations of the sample of Multi-planet Systems were supported by Keck time from the University of California.  The samples of Ultra-Short Period Planets and Habitable Zone Planets were observed using Keck time from NASA and the California Institute of Technology specifically for this project.  Most of the CKS results ($\sim$1000 stars) are derived from spectra reported for the first time here.  Some of the CKS stars ($\sim$300/\nstars) were observed with Keck-HIRES as part of the NASA Keck time awarded to the KFOP team specifically for mission support and are included in CKS. 
Those previous observations were for characterization of noteworthy systems or as part of determining precise planet masses. The KFOP observations are described in \Kepler Data Release 25 \citep[DR25;][]{Mathur2016} and include spectroscopic parameters that may vary slightly compared with our results. 
See Furlan et al., in prep.\ for a summary of KFOP spectroscopy.
All spectra used in this paper are publicly available on Keck Observatory Archive.

\subsection{Observations}
We observed all \nstars stars in the CKS sample with the HIRES spectrometer \citep{Vogt1994} at the  W.\ M.\ Keck Observatory. We used an exposure meter to stop the exposures after achieving a signal-to-noise ratio (S/N) of 45 per pixel (90 per resolution element) at the peak of the blaze function in the spectral order containing 550 nm.  A small subset of targets was observed at higher S/N, usually because a higher S/N was needed to serve as template spectra for precise RV measurements \citep{Marcy2014}. For the faintest targets ($Kp>15.0$) the S/N was limited to $\sim$20 per pixel, given the constraints on the total observing time. The spectral format and HIRES settings were identical to those used by the California Planet Search \citep{Howard2010_CPS}. This includes the use of the B5/C2 decker with dimensions of $0\farcs86 \times 3\farcs5 / 0\farcs86 \times 14\farcs0$, resulting in a spectral resolution of 60,000. For stars with $V > 11$ (most of the sample), we used the C2 decker and employed a sky-subtraction routine to reduce the impact of scattered moonlight and telluric emission lines \citep{Batalha2011}.  The spectral coverage extended from 3640 to 7990~\AA. We aligned the spectral format of HIRES such that the observatory-frame wavelengths were consistent to within one pixel from night to night.  This allows for extraction of the spectral orders using the CPS raw reduction pipeline. We used the HIRES guide camera with a green filter (BG38), ensuring that the guiding signal was based on light near the middle of the wavelength range of the spectra. Except for a few stars with nearby companions, we used the HIRES image-rotator in the vertical-angle mode to capture the full spectral bandwidth within the spectrometer entrance slit.

\subsection{Stellar Samples}
\label{sec:stellar_samples}
\begin{figure}
\epsscale{1.0}
\plotone{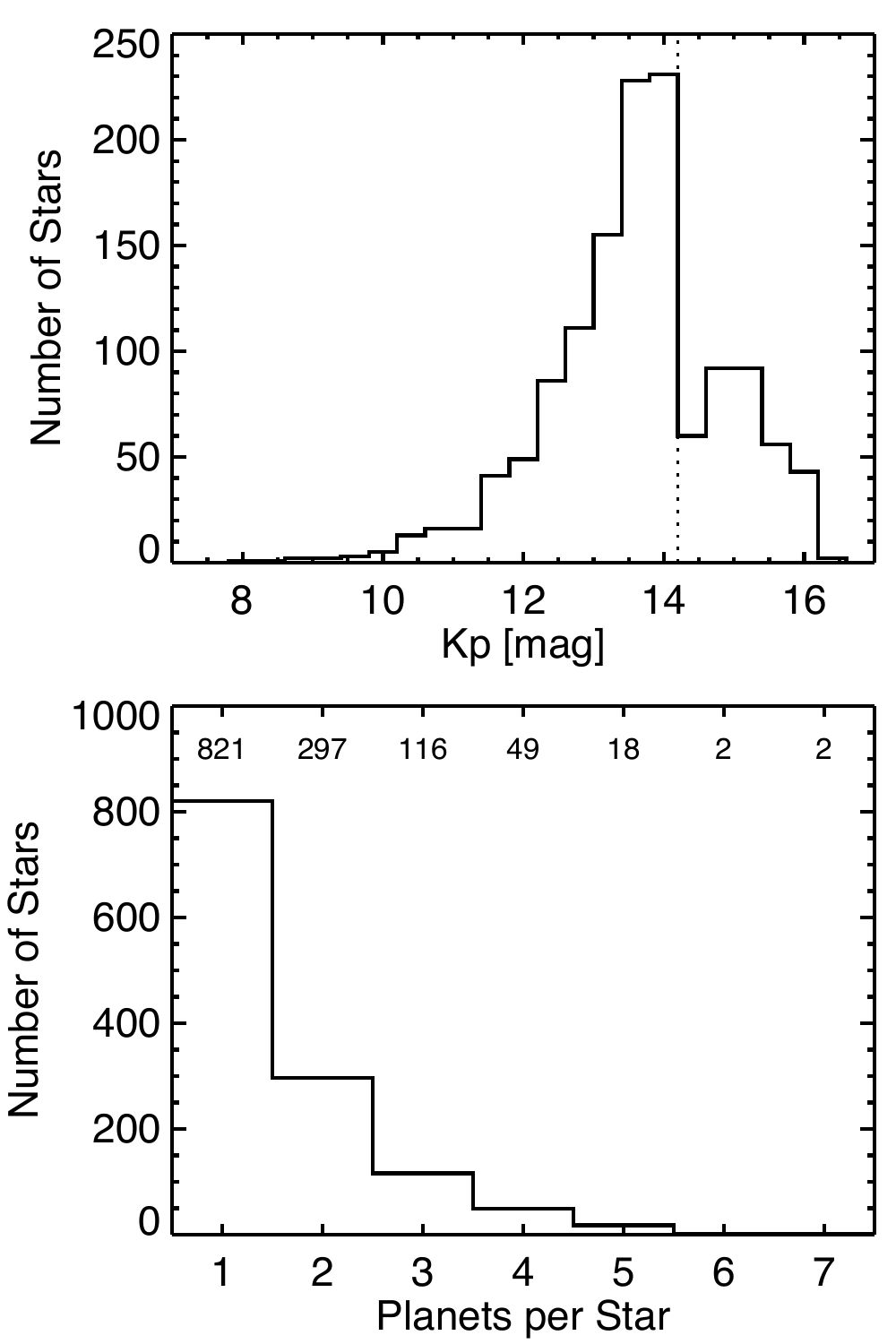}
\caption{Properties of the CKS sample.  Top: Distribution of stellar brightness in the \Kepler bandpass (\Kp).  The dashed line at \Kp\ = 14.2 indicates that faint limit of the magnitude-limited sample.  Bottom: Distribution of the number of planets per star.  The label above each histogram bin specifies the number of stars belonging to that bin.}
\label{fig:kepmag_dist}
\end{figure}

The CKS sample comprises several overlapping sub-samples listed below. Table~\ref{tab:samples} provides a summary of the number of stars and planets belonging to each subsample while Table~\ref{tab:target_list} provides the star-by-star designations. Figure~\ref{fig:kepmag_dist} shows the distribution of stellar brightness and of the number of planets per star, for the entire CKS sample.

\begin{deluxetable}{lrr}
\tabletypesize{\footnotesize}
\tablecaption{CKS Stellar Samples\label{tab:samples}}
\tablewidth{0pt}
\tablehead{
  \colhead{Sample}   & 
  \colhead{$N_\mathrm{stars}$}   &
  \colhead{$N_\mathrm{planets}$}
}
\startdata
Magnitude-limited ($Kp<14.2$) & \nstarsMaglim & \nplanetsMaglim\\
Multi-planet Systems & \nstarsMulti & \nplanetsMulti\\
Habitable Zone Systems & \nstarsHZ & \nplanetsHZ\\
Ultra-Short Period Planets  & \nstarsUSP & \nplanetsUSP \\
Other & \nstarsOther & \nplanetsOther\\
False Positives\tablenotemark{a}  & \nstarsFP & \nplanetsFP \\
Total\tablenotemark{b} & \nstars & \nplanets
\enddata
\tablenotetext{a}{The False Positive sample includes systems for which {\it all} planet candidates have been dispositioned as false positives.}
\tablenotetext{b}{Some stars are in multiple samples.}
\end{deluxetable}

\textit{Magnitude-limited.}  This sample is defined as all stars with $Kp< 14.2$. We set out to observe a magnitude-limited sample of KOIs chosen independent of the number of detected planets or previously measured stellar properties. As the project progressed, we added additional samples of fainter stars, as described below.

\textit{Multi-planet Systems.}  This sample is defined as KOIs stars orbited by two or more \textit{transiting} planets (excluding false positives).  We also observed nearly all of the multi-transiting systems appearing in the \cite{Rowe2014} catalog, with priority given to the highest multiplicity systems and the brightest stars.  CKS Paper V (Weiss et al.\ in prep.) performs a detailed analysis of the multi-planet systems.

\textit{Habitable-Zone Systems.}  We observed \nstarsHZ host stars of \Kepler planets residing in or near the habitable zone defined by  \citep{Kopparapu2013}. Some of the individual habitable-zone planets have been studied extensively and validated \citep{Borucki2013,Torres2015,Jenkins2015}.  
It is not clear what to adopt as the boundaries of the liquid-water habitable zone, because of the many uncertainties in exoplanet atmospheric properties and other factors that impact planet habitability \citep{Seager2013}. 
The NASA \Kepler Team constructed a list of habitable-zone targets using the best available stellar parameters at the time.  They selected stars for which the flux received by the planet fell (within 1$\sigma$) between the Venus and ``early-Mars'' habitable-zone boundaries \citep{Kopparapu2013}. 
After the revision to the stellar parameters based on our CKS spectra, we now know that some of these planets are well outside of the habitable zone.  
CKS Paper II (Johnson et al., submitted) gives the newly determined values for stellar flux and planetary equilibrium temperature for all the CKS stars.

\textit{Ultra-Short Period Planets.}  Ultra-short period (USP) planets \citep{Sanchis-Ojeda2014} have orbital periods shorter than one day. Winn et al.\ (2017, submitted) have performed an investigation of this sample, in particular on the metallicity distribution.

\textit{Other.} We observed 38 additional \Kepler planet host stars for reasons that do not fall into any of the preceding categories.  Often these {\it ad hoc} observations were for studies of unusual or noteworthy planetary systems (e.g. \citealt{Dawson2015,Desert2015,Holczer2015,Kruse2014}).

\textit{False Positives.}  The planetary candidate status (``disposition'') of some KOIs has changed over time. Inevitably we observed KOIs that are now recognized as false positives.  For completeness we report on the parameters for these false positives. Importantly, though, the false positives were not used for the cross-calibration between our two spectroscopic analysis pipelines (see Sec.~\ref{sec:calibration-sme}). More details on this sample are given in
Sec.\ \ref{sec:false_positives}.


It is important to recognize that the samples in the CKS survey are built upon the foundation of the \Kepler mission.
Assembling the \Kepler planet catalogs required the extraordinary effort and devotion of the \Kepler team members \citep{Borucki2011,Batalha2013, Burke2014,Rowe2015}. Also essential was the painstaking engineering behind the photometer \citep{Caldwell2010,Gilliland2011,Bryson2010,Haas2010}, as well as the software engineering that transformed CCD pixel values into planet candidates \citep{Jenkins2010,Gilliland2010,Stumpe2012,Smith2012,Smith2016,Batalha2010,Batalha2010b,Torres2011,Bryson2013,Christiansen2012,Christiansen2013,Christiansen2015,Christiansen2016,Thompson2015,McCauliff2015,Tenenbaum2013,Tenenbaum2014,Twicken2016,Kinemuchi2012}. 

\begin{deluxetable*}{lcccccc}[tb]
\tabletypesize{\footnotesize}
\tablecaption{CKS Target Stars\label{tab:target_list}}
\tablewidth{0pt}
\tablehead{
  \colhead{} & \multicolumn{6}{c}{Stellar Samples}  \\ 
  \cline{2-6}   \\
  \colhead{}   & 
  \colhead{Magnitude-limited}   &
  \colhead{Multi-planet}   &
  \colhead{Habitable}   &
  \colhead{Ultra-Short}   &
  \colhead{}   &
  \colhead{All Planets are}\\
  \colhead{KOI No.}   & 
  \colhead{(\Kp\ $<$ 14.2)}   &
  \colhead{Systems}   &
  \colhead{Zone}   &
  \colhead{Period Planets}   &
  \colhead{Other}   &
  \colhead{False Positives}
}
\startdata
          1 &   1 &   0 &   0 &   0 &   0 &   0 \\
          2 &   1 &   0 &   0 &   0 &   0 &   0 \\
          3 &   1 &   0 &   0 &   0 &   0 &   0 \\
          6 &   1 &   0 &   0 &   0 &   0 &   1 \\
          7 &   1 &   0 &   0 &   0 &   0 &   0 \\
\tablecomments{This table will be published in its entirety in the machine-readable format in the accepted version of this paper. A portion is shown here for guidance regarding its form and content.}
\tablenotetext{}{Stars marked ``1'' are members of a stellar sample while those marked ``0'' are not.}
\end{deluxetable*}

\begin{deluxetable}{lcccc}[tb]
\tabletypesize{\footnotesize}
\tablecaption{CKS Candidate Planets \label{tab:disposition_table}}
\tablewidth{0pt}
\tablehead{
  \colhead{} & \colhead{} & \multicolumn{3}{c}{False Positive Assessment}  \\ 
  \cline{3-5}   \\
  \colhead{KOI}   & 
  \colhead{Adopted}   &
  \colhead{Morton\tablenotemark{b}}   &
  \colhead{Mullaly\tablenotemark{c}}   &
  \colhead{NEA\tablenotemark{d}}   \\
  \colhead{Candidate}   & 
  \colhead{Disposition\tablenotemark{a}}   &
  \colhead{}   &
  \colhead{}   &
  \colhead{}   
}
\startdata
K00001.01 & CP & CP & CP & CP  \\ 
K00002.01 & CP & CP & CP & CP  \\ 
K00003.01 & CP & CP & CP & CP  \\ 
K00006.01 & FP & FP & FP & FP  \\ 
K00007.01 & CP & CP & CP & CP  \\ 

\enddata
\tablecomments{This table will be published in its entirety in the machine-readable format in the accepted version of this paper. A portion is shown here for guidance regarding its form and content.}
\tablenotetext{a}{Dispositions: CP = confirmed planet; PC = planet candidate; FP = false positive.}
\tablenotetext{b}{\cite{Morton2016} }
\tablenotetext{c}{\cite{Mullally2015}}
\tablenotetext{d}{NASA Exoplanet Archive, accessed 2017 February 1; \url{http://exoplanetarchive.ipac.caltech.edu}}
\end{deluxetable}

\subsection{Spectral Archive}
\label{sec:spectral_archive}
All stellar spectra analyzed here are available to the public via the Keck Observatory Archive,%
\footnote{\url{http://www2.keck.hawaii.edu/koa/public/koa.php}}
the Community Follow-up Program (CFOP) website,%
\footnote{\url{http://cfop.ipac.caltech.edu}}
and the CKS project website.%
\footnote{\url{http://astro.caltech.edu/~howard/cks/}}
The CFOP website also contains additional information about each KOI and a discussion of the available follow-up observations.  We have also made available the standard rest-frame wavelength solution applicable to every spectrum, which is accurate to within one pixel.
 
\begin{figure}
\includegraphics[width=0.85\linewidth]{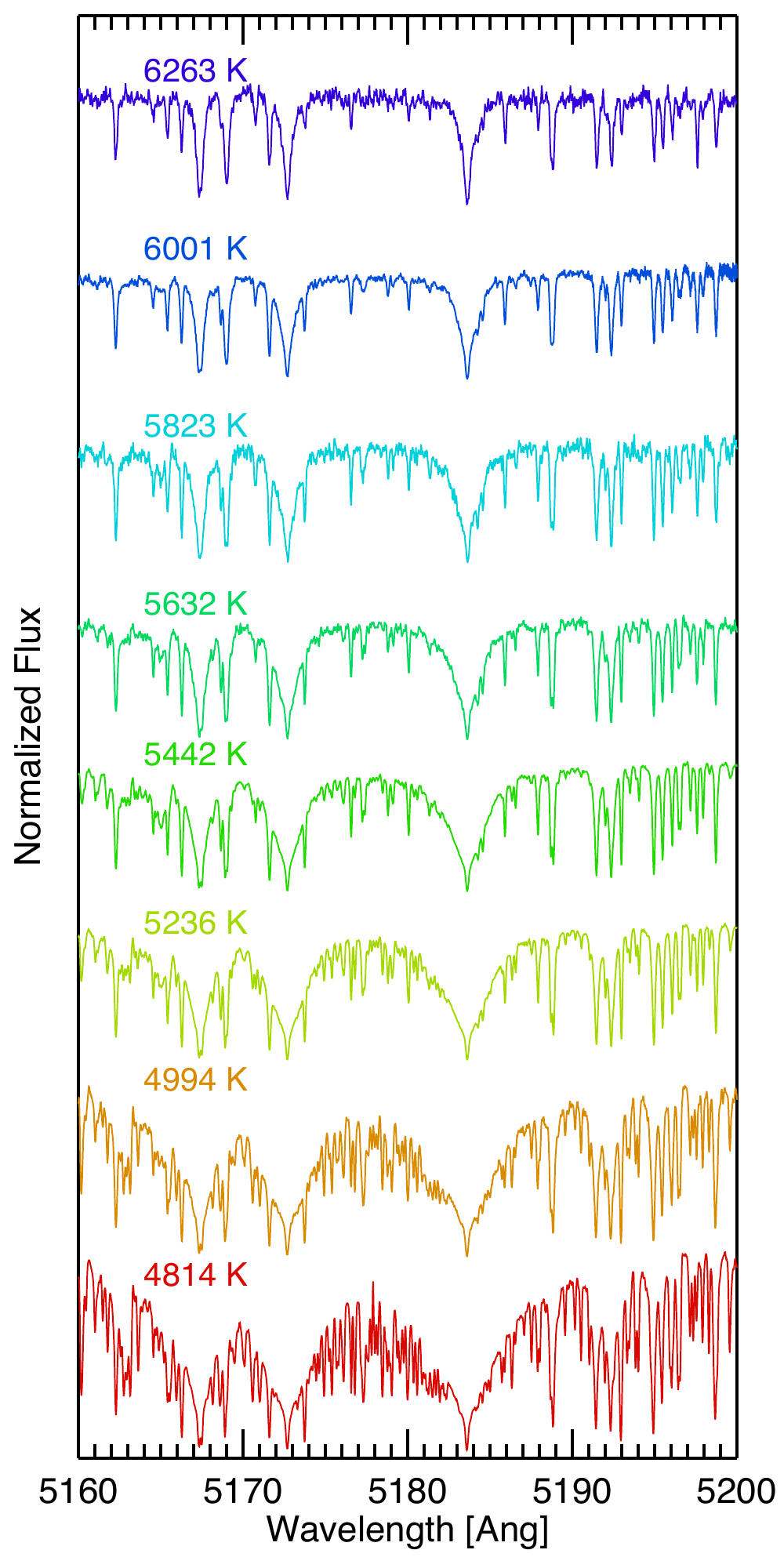}
\centering
\caption{Keck-HIRES spectra spanning the Mg I b lines of eight slowly-rotating main sequence CKS stars, in $\sim$200 K increments of effective temperature. }
\label{fig:stellar_spectra}
\end{figure}

One auxiliary data product is the measurement of each star's velocity relative to the Solar System barycenter, as determined from measurements of the telluric absorption features \citep{Chubak2012}. 
These systemic radial velocities have a precision of 0.1~\kms and are listed along with the spectroscopic parameters. 
Other auxiliary products are the stellar activity indicators that fall onto the HIRES format. The \caii\ measurements for this sample, in conjunction with stellar photometry, will be valuable when determining age-activity-rotation correlations \citep{Isaacson2010}. Figure\ \ref{fig:stellar_spectra} shows some typical CKS spectra containing the Mg I b region for a range of effective temperatures along the main sequence.
In addition, \cite{Kolbl2015} consolidates all the available identifications of secondary spectral lines due to a second star that was admitted into the spectrometer slit.

\subsection{False Positive Identification}
\label{sec:false_positives}

``False positives'' are KOIs that were initially classified as planet candidates, but later deemed to be non-planetary in nature. The most common types of false positives are foreground eclipsing binaries, background eclipsing binaries, and data artifacts. Statistical considerations of false positive scenarios suggest that false positives account for $\sim$10\% of all the planet candidates \citep{Morton2011,Morton2012, Fressin2013}. KOIs hosting multiple planet candidates have an even lower false positive contamination rate of $\sim1$\% \citep{Lissauer_2012,Lissauer2014,Rowe2014}. In contrast, \cite{Santerne2012} found a higher false positive contamination rate among gaint planet candidates of 30\% through radial velocity follow-up.

False positives due to data artifacts can be caused by reflections from \Kepler's primary mirror and spillover light from eclipsing binaries that occupy nearby pixels on the \Kepler CCD. Identifying false positives by matching KOI ephemerides to known eclipsing binaries revealed several hundred false positives and further improving the quality of later \Kepler candidate lists \citep{Coughlin_2014}.


We identified false positives in our sample by cross-matching with published false positive catalogs and on-line resources.  In addition to planet catalogs produced by the \Kepler team, detailed planet validation has been performed by \cite{Lissauer_2012,Lissauer2014} on mulit-planet systems, and on large samples (1000's) of KOIs by \cite{Mullally2015} and \cite{Morton2016}. The NASA Exoplanet Archive \citep{Akeson2013} hosts a cumulative list of dispositions for every KOI.  All of these catalogs combined provide high quality vetting of the \Kepler planet candidate lists. Follow-up observations by the KFOP and community at large, using ground based facilities have also contributed heavily to false positive analysis as well as stellar classification, which has been used to improve the integrity of the planet candidate and confirmed planet lists.

For this work, we assign dispositions to the KOIs by referring to three catalogs. For each KOI, we first consult the catalog of \cite{Morton2016} and adopt that catalog's disposition whenever it is available. If the KOI does not appear in that catalog, we seek a disposition in the catalog of \cite{Mullally2015}. If neither of those catalogs gives a disposition, we adopt the disposition of the NASA Exoplanet Archive. Our catalog does not contain any cases for which the KOI has conflicting dispositions of ``false positive'' and ``confirmed-planet/planet-candidate''. All the KOIs in our sample are either confirmed planets, planet candidates, or false positives. Table \ref{tab:disposition_table} gives the dispositions that follow from this procedure, and that are adopted for this and the subsequent CKS papers.

Upon closer examination of several KOIs for which our spectroscopic analysis produced suspect results, we identified 8 KOIs as false positives. Several are eclipsing binaries (KOIs 113, 134, 1032, 1463, 3419) and one is a brown dwarf (KOI-415) as determined with radial velocity measurements by \cite{Moutou2013}. The transit signal detected in KOI-1546 was shown to arise from the variations of a different star in the field.  KOI-1739 is a single-lined spectroscopic binary as determined via radial velocity measurements. The status of the latter two KOIs is documented on the CFOP. Because our spectroscopic pipelines assume a single spectrum, we removed the double-lined stars identified by \cite{Kolbl2015} from the CKS sample; the characteristics of those systems can be found in Table 9 of \cite{Kolbl2015}. 

\section{Spectroscopic Pipelines}
\label{sec:pipelines}
We measured the stellar spectroscopic parameters using two independent data analysis pipelines: \SpecMatch and \SME. We describe these two pipelines in Sec.~\ref{sec:specmatch} and \ref{sec:sme}, respectively. The two separate techniques permit the identification of suspect spectroscopic parameters by looking for large inconsistencies between the two methods. We describe the two analysis methods below.  Sec.~\ref{sec:catalog} gives the details of the construction of the combined catalog of stellar parameters.

\subsection{SpecMatch}
\label{sec:specmatch}
\SpecMatch is a publicly-available%
\footnote{https://github.com/petigura/specmatch-syn}

tool for precision stellar characterization, developed specifically for the CKS project, to accommodate spectra with a lower signal-to-noise ratio (S/N) than the usual spectra processed by the California Planet Search (CPS) team. Precision stellar characterization using HIRES spectra has been a key component in the exoplanet work of CPS for two decades. Typically, such analyses are performed using high signal-to-noise ``template'' observations, obtained during the course of the team's RV observations \cite{Marcy1992}. These template spectra typically have a S/N of 150--200 per HIRES pixel, permitting detailed modeling of several narrow regions of the spectrum with realistic stellar atmosphere models \citep{Valenti05}. To compensate for the lower S/N of the CKS spectra, \SpecMatch fits $\approx$400 \AA\ of the spectrum using computationally-efficient interpolation between precomputed model spectra, as opposed to detailed spectral synthesis.

Here, we offer a brief summary of the \SpecMatch algorithm; for further details see \cite{Petigura2015}. \SpecMatch fits five segments of an observed optical spectrum using forward-modeling. The code creates a synthetic spectrum of arbitrary \teff, \logg, \feh, and \vsini by first interpolating between model spectra computed by \cite{Coelho2005} at discrete values of \teff, \logg, and \feh. Next, \SpecMatch accounts for line broadening due to stellar rotation and convective macroturbulence by convolving the interpolated spectrum with the kernel specified by \cite{Hirano11}. Then, \SpecMatch accounts for the instrumental profile of HIRES, which we model as a Gaussian having a FWHM of 3.8~HIRES pixels. We choose this value because it can reproduce the width of telluric lines observed through the ``C2'' decker for typical seeing conditions (see \citealt{Petigura2015} for further details). The version of \SpecMatch used in this work has been slightly modified from the version presented in \cite{Petigura2015}. Instead of modeling all five spectral segments simultaneously, we model each segment individually and average the resulting parameters at the end. This modification improved run time, and the consistency of the parameters derived from individual segments provides a good check on the quality of the \SpecMatch fits. 

\cite{Petigura2015} verified the precision and accuracy of \SpecMatch parameters by comparisons with well-characterized touchstone stars from the literature. After calibrating the gravities to asteroseismic values computed by \cite{Huber_2013}, \cite{Petigura2015} found that \SpecMatch reproduces the surface gravities determined through asteroseismology to within 0.08~dex (RMS). \cite{Petigura2015} demonstrated a precision in effective temperature and metallicity of 60~K and 0.04~dex, respectively, based on comparisons with \cite{Valenti05}. Finally, \cite{Petigura2015} demonstrated a precision in projected stellar rotation, \vsini, of 1.0~\kms, for $\vsini \geq 2.0$~\kms.

Calibrating the \SpecMatch \logg values to the \cite{Huber_2013} scale has the following shortcoming: the calibration is only valid over the domain of the HR-diagram containing stars with asteroseismic measurements, i.e. evolved stars and main sequence stars having spectral type $\sim$G2 and earlier. Extending the calibration toward later spectral types is a risky extrapolation, and reverting to the uncalibrated \SpecMatch parameters introduces a discontinuous correction. Recently, \cite{Brewer16} (B16 hereafter) extended the work of \cite{Valenti05} by performing a detailed spectroscopic analysis of 1617 CPS target stars with updated version of SME \citep{Brewer15}. The B16 catalog is an ideal calibration sample for \SpecMatch because the spectroscopic surface gravities reproduce asteroseismic surface gravities to 0.05~dex and there is a large overlap in stars analyzed by both techniques.

\begin{figure*}
\centering
\hspace*{1cm}    
\includegraphics[width=0.95\textwidth]{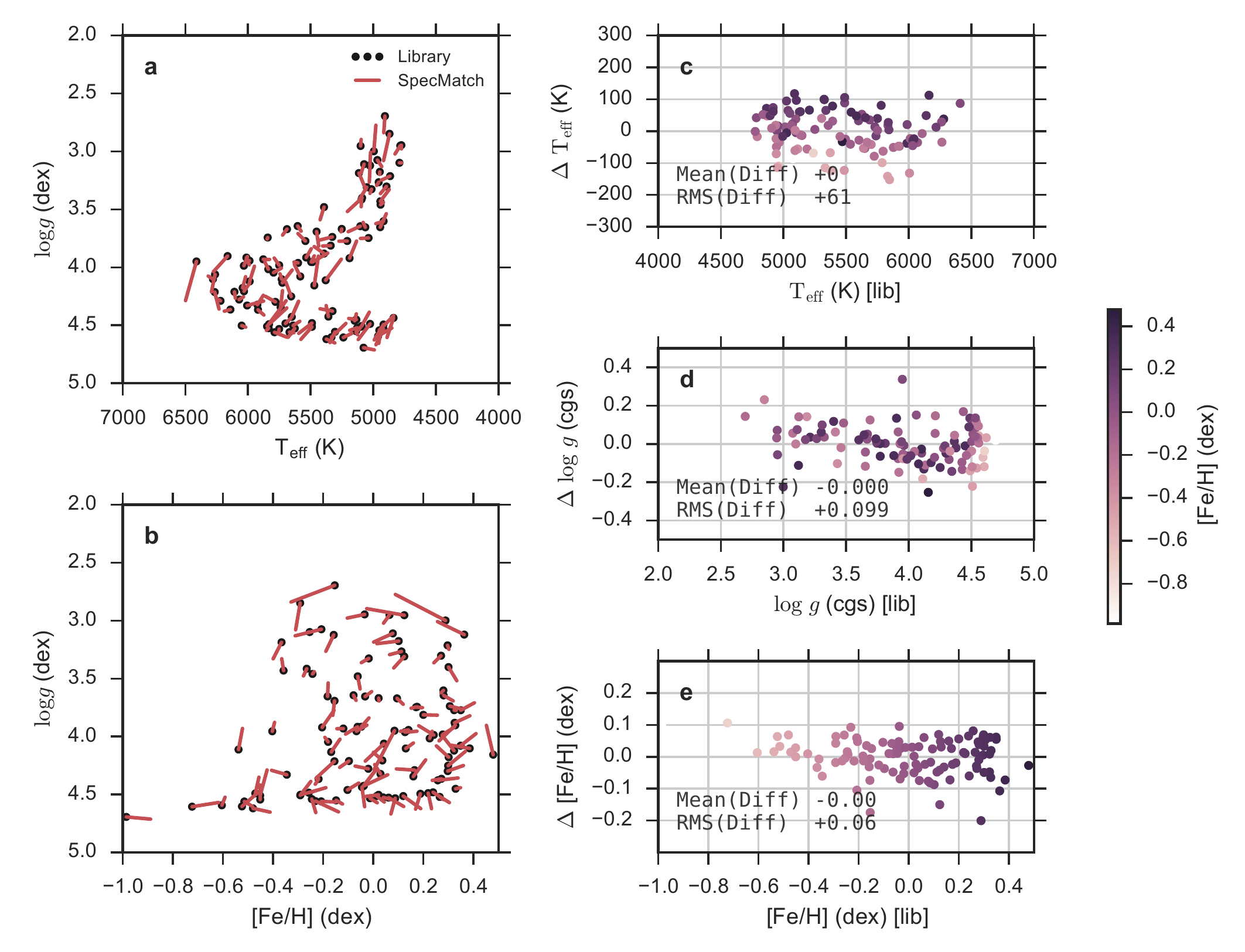}
\caption{Comparison of stellar parameters from the \cite{Brewer16} (B16) spectroscopic analysis and \SpecMatch. {\bf a)} Black points show \teff and \logg from B16 and red lines point to the \SpecMatch values. Shorter lines correspond to tighter agreement. {\bf b)} Same as {\bf a)}, except showing log g and [Fe/H]. {\bf c)} Differences in \teff between \SpecMatch and B16, i.e. $\Delta \teff$ = \teff (SM) $-$ \teff (B16), as a function of \teff (B16). Points are colored according to B16 metallicity {\bf d-e)} Same as {\bf c)}, except showing \logg and \feh, respectively. Dispersion (RMS) in $\Delta \teff$, $\Delta \logg$, $\Delta \feh$ is 61 K, 0.099 (dex), 0.06 (dex), respectively. We note a residual correlation between $\Delta \teff$ and \feh in {\bf c)} of $\approx$10~K per 0.1~dex. For the sake of simplicity, we elected not to calibrate out this trend. The systematic is reflected in the 60 K (RMS) scatter in $\Delta \teff$ and in our adopted uncertainties of \cksSigTeff~K.}
\label{fig:brewer-cal-1-on-brewer-cal-1-cal}
\end{figure*}

We calibrated the \SpecMatch parameters to the B16 scale by selecting 106 from the 1617 stars analyzed by B16 that spanned the following range of parameters: $\teff = 4700-6500$~K, $\logg = 2.50-4.75$~dex, and $\feh = -1.0-+0.5$~dex. For each parameter, we derived a correction $\Delta$ that calibrates the \SpecMatch parameters onto the B16 scale via $\mathrm{SM}_\mathrm{cal} = \mathrm{SM}_\mathrm{raw} + \Delta$. The corrections are linear (and therefore continuous) functions of the following form:
\begin{eqnarray}
\Delta \teff & = & a_0 + a_1 \left(\frac{\teff - 5500~\K}{100~\K}\right), \nonumber\\
\Delta \logg & = & b_0 + b_1 \left(\frac{\logg - 3.5~\dex}{0.1~\dex}\right) +  b_2 \left(\frac{\feh}{0.1~\dex}\right), \nonumber\\
\Delta \feh  & = & c_0 + c_1 \left(\frac{\feh}{0.1~\dex}\right), \nonumber
\end{eqnarray}
where $a_0 = -61.9~\K$ and $a_1  = 6.13$; $b_0 = -0.0234~\dex$, $b_1 = -0.0026$; and $b_2 = -0.0412$, $c_0 = 0.0150~\dex$, and $c_1 = -0.0126$. The coefficients were chosen such that they minimized the RMS difference between the calibrated \SpecMatch and B16 parameters (i.e. $\mathrm{B16} - \mathrm{SM}_\mathrm{cal}$).  After applying these corrections, we compare the calibrated \SpecMatch and B16 parameters in Figure~\ref{fig:brewer-cal-1-on-brewer-cal-1-cal}. We find a dispersion of \BrewerTeffRMS~K, \BrewerLoggRMS~dex, and \BrewerFehRMS~dex in \teff, \logg, and \feh, respectively. By construction, there is no mean offset between the calibrated \SpecMatch and B16 parameters. We verified that the flexibility in our calibration was not misrepresenting the agreement between \SpecMatch and B16 by comparing a distinct group of 80 stars that were not used in the calibration. The agreement between B16 and \SpecMatch was comparable for this second set of stars: RMS dispersions were 55~K, 0.10~dex, and 0.05~dex for \teff, \logg, and \feh, respectively, and mean offsets were small at 5~K, 0.00~dex, and 0.00~dex, respectively. We refer to the calibrated \SpecMatch parameters hereafter.

\subsection{\SME}
\label{sec:sme}

We also measured \teff, \logg, \feh, and \vsini using \SME, a set of Python routines wrapped around the widely-used spectral synthesis program, Spectroscopy Made Easy \citep[SME;][]{Valenti1996}. Stellar characterization with SME is done with the spectral synthesis technique which generates a synthetic spectrum that matches the observed data by performing radiative transfer through a model atmosphere based on a set of global stellar properties.   \SME automates the process of spectral synthesis, facilitating the analysis of large data sets of high-resolution spectra in 
order to determine robust stellar parameters with realistic uncertainties in a hands-off fashion.

At its core, \SME uses version 342 of the SME program.  SME~342 has three main components: 
the radiative transfer engine, the software that interpolates the grid of model atmospheres, and the 
Levenberg-Marquardt non-linear least-squares solver that finds the optimal solution.
As the SME~342 solver converges from the initial guesses to a set of best-fit free parameters, each step in the $\chi^2$ 
minimization process requires an interpolation of the input model atmosphere grid at the specific set of global 
parameters, and then a new solution is found of the radiative transfer equations through this specific model atmosphere.  
SME~342 uses a fast radiative transfer algorithm based on the SYNTH code \citep{Piskunov1992}. This employs
an adaptive wavelength grid, in which the 
density of radiative transfer calculations is adjusted to increase the spectral resolution in the vicinity of absorption lines and decrease the resolution 
in regions of the continuum.
The structure of stellar atmospheres includes steep gradients with curvature of density, pressure, and temperature, therefore, SME~342 uses a specialized routine to perform non-linear Bezier interpolation of a grid of atmosphere models in order to predict a stellar atmosphere at a specific set of global stellar parameters.

\SME requires as input (1) a set of plane-parallel model atmospheres, (2) a list of atomic and molecular lines and their associated line parameters (i.e.\ a line list), and (3) initial guesses for the free parameters.
When analyzing the CKS stars, \SME ingests a grid of plane-parallel MARCS model atmospheres \citep{Gustafsson2008} calculated under conditions of local thermodynamic equilibrium and spanning the range of potential stellar parameters in \teff, \logg, and \feh.
Since the model atmospheres have plane-parallel geometry and do not include a realistic treatment of convection, SME introduces 
the microturbulent and macroturbulent velocity parameters ($V_{\rm mic}$ and $V_{\rm mac}$, respectively) to achieve better agreement between 
the synthetic and observed spectra.   In \SME, we adopt empirical analytic functions for the behavior of the 
micro- and macroturbulent velocities that are dependent on \teff.  
Specifically, we use a relationship for the microturbulent velocity given by \cite{Gomez2013}. For the macroturbulent velocity, we incorporate the relationship given by \cite{Valenti05}.%
\footnote{With the sign correction specified in Footnote 6 of \citep{Torres2012}}
We also note that instead of using one fixed velocity throughout the $\chi^2$ minimization, we use dynamic values that are adjusted appropriately at each minimization step based on the effective temperature.

\SME uses a line list and abundance pattern adapted from \citet{Stempels2007} and \citet{Hebb2009}. This line list contains atomic and molecular transition information taken from the VALD database and the information provided on Robert Kurucz's website \citep{Kurucz1975}. The wavelengths included in the spectral synthesis are the region around the Mg~b triplet (5150--5200~\AA), the NaI~D doublet region (5850--5950~\AA), and the wavelength region of 6000--6200~\AA\ which contains many isolated atomic lines and is relatively free of telluric features. We have incorporated the empirical corrections to the oscillator strengths and broadening parameters for individual lines determined by \citet{Stempels2007} through a comparison between a high-resolution spectrum of the Sun \citep{Kurucz1984} and a synthetic spectrum calculated using the spectroscopic parameters of the Sun.

\begin{figure}
\includegraphics[width=0.48\textwidth]{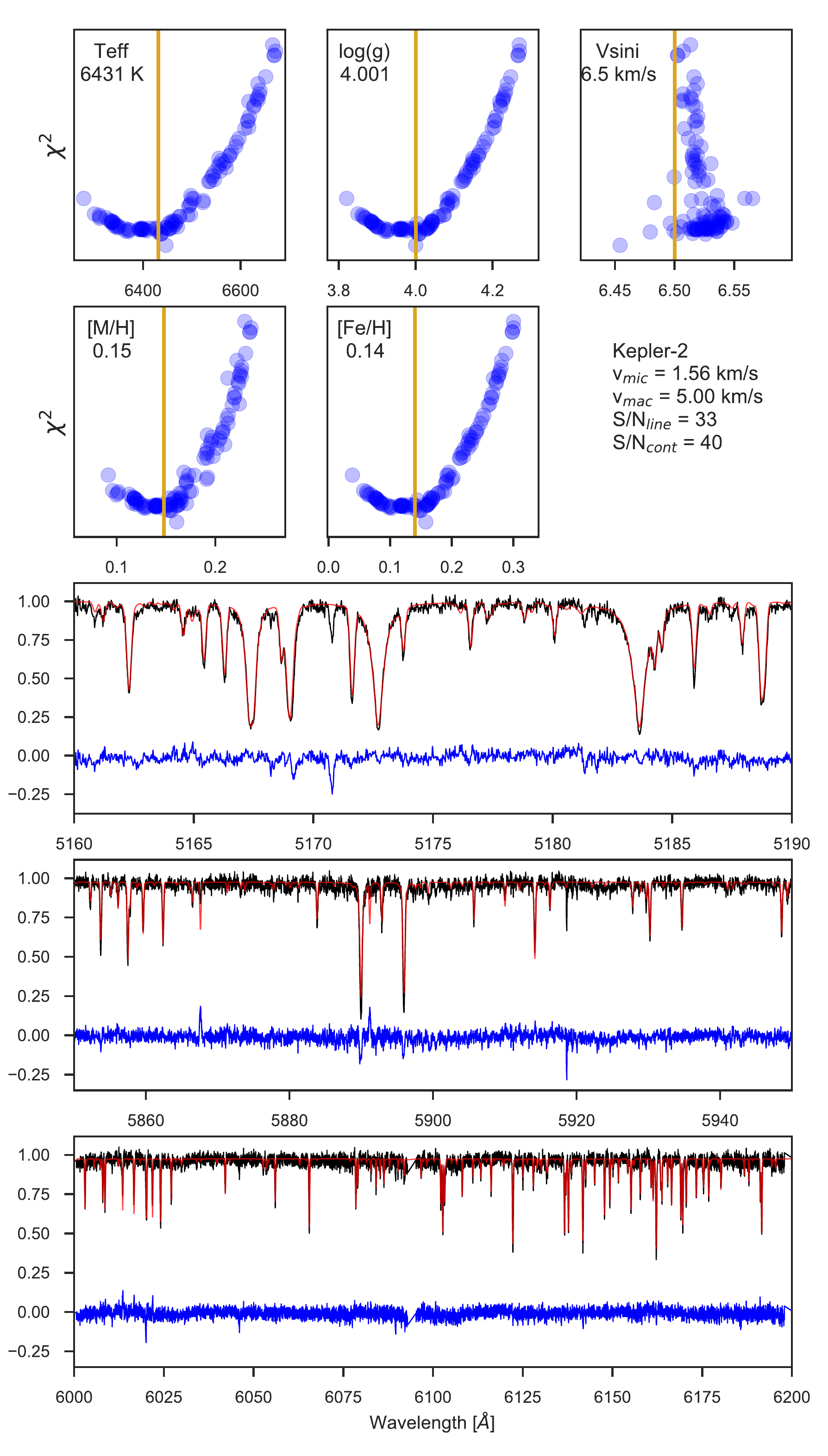}
\caption{Output from \SME for the CKS spectrum of Kepler-2. Top panels show for each global parameter the output distribution of $\chi^2$ values for a set of initial guesses. The vertical lines show the determined best-fit parameters. The bottom three panels show observed spectrum (black), synthesized spectrum (red), and residuals (blue).}
\label{fig:sme}
\end{figure}

Like any Levenberg-Marquardt based solver, SME~342 requires a good initial guess
and a smoothly varying $\chi^2$ surface in order to consistently find the optimal solution at the absolute global minimum.  Unfortunately, the discreteness of the wavelength and stellar atmosphere grids utilized by SME~342 add artificial structure to the $\chi^2$ surface and hinder convergence. In addition, without {\em a~priori} information about the free parameters, a single run of SME~342 can become stuck in a local minimum and fail to converge to the global solution. Historically, the $\chi^2$ minimum has been  found through hands-on manipulation by an expert SME user. \SME solves this problem and automates the spectral synthesis process by running many realizations of SME~342 starting from different initial conditions.  Due to the convergence issues with a single run of SME~342, the distribution of output solutions from the multiple trials performed by \SME results in a sampling of the $\chi^{2}$ surface close to the global minimum which \SME uses to identify the best stellar parameters and their uncertainties (see Figure~\ref{fig:sme}) .

Using this approach, we analyzed \nstarsSX/\nstars CKS spectra on the Stampede computer cluster at NSF's XSEDE facility.  (A few stars were not analyzed by \SME because their spectra were not gathered when the computing time was available.  \SpecMatch parameters are available for those stars.)
The automated \SME run on each star includes 98 realizations of SME~342 started from a range of different initial conditions determined by randomly drawing from uniform distributions around the {\it Kepler} Input Catalog parameters for each star.
After the initial \SME run is complete, a further check is performed to insure that the distributions of output values is smaller and fully encompassed by the range of initial guesses. If not, \SME was re-run with a wider range of initial guesses. This limits bias in final stellar parameters due to selecting only a small or systematically skewed set of initial guesses.

Figure~\ref{fig:sme} shows the output from an \SME analysis of Kepler-2 \citep[also known as HAT-P-7;][]{Pal2008}.  The top five panels show the final $\chi^2$ distribution versus each of the free parameters (\teff, \logg, \feh, \vsini, and [M/H]) resulting from the 98 independent realizations of SME~342.  The majority of runs do not converge to the global $\chi^2$ minimum, but the resulting distribution of final values describes the $\chi^2$ surface which we use to determine the optimal parameter values at the global minimum (dark tan line) and the asymmetric $1\sigma$ uncertainties on these values (light tan region).   The bottom three panels show the observed spectrum in the synthesis regions (black) with the best fitting synthetic spectrum over-plotted (red).    

\section{Catalog of Stellar Properties}
\label{sec:catalog}
We compared the outputs of the two different codes, \SpecMatch and \SME. For our final results, we combined the parameters produced by both codes, after making small adjustments to the raw \SME values to place them on the SpecMatch scale. Figure \ref{fig:cks-dumbbell-steff-smet} shows the spectroscopic HR diagram (\teff, \logg) for the \SpecMatch, \SME, and combined catalogs. Figures \ref{fig:cks-dumbbell-steff-smet} and \ref{fig:cks-dumbbell-smet-slogg} show projections of the parameters into the (\teff, \feh) and (\feh, \logg) planes. We describe our procedure for combining the two spectroscopic catalogs below.

\subsection{SpecMatch and \SME Catalogs}
The catalogs of stellar properties produced by \SpecMatch and \SME show excellent agreement in \teff, \logg, and \feh for stars that are not deemed false positives.
Figure~\ref{fig:sx-on-sm} shows the differences between the raw results from the two pipelines, analyzing the same stellar spectrum. The systematic differences in \teff, \logg, and \feh determinations are small, with median and RMS differences between \SpecMatch and \SME analyses of the same stellar spectra being comparable to the individual measurement errors. We correct for the small systematic differences as described below.  The independent \vsini measurements do not agree, however. This is due to the use of a Gaussian instrumental profile with a resolution of $R\sim$ 75,000 in \SME, which is higher than the empirically-determined value used in \SpecMatch.  Because of this known systematic issue in the \vsini values from \SME, we adopt the \SpecMatch values of \vsini for all stars in this catalog.

\subsection{Calibration of \SME Parameters}
\label{sec:calibration-sme}
We attempted to correct the low-order differences between the \SpecMatch and \SME parameters to put the catalogs on the same scale. We adopted \SpecMatch as the standard since it is well-calibrated for all parameters. In addition, a comparison of the \vsini\ values for 43 stars with Rossiter-McLaughlin of transiting giant planets \citep{Albrecht2012} and \SpecMatch analyses showed agreement at the level of 1 \kms (RMS). Because of this heritage, we made minor corrections to the \SME parameter values and left the \SpecMatch values unchanged.  

We compare the differences between the \SpecMatch and \SME parameters in Figure~\ref{fig:sx-on-sm}. Following the methodology of Section~\ref{sec:specmatch}, we derived a correction that calibrates the \SME parameters onto the \SpecMatch (and B16 scale) via $\mathrm{SX}_\mathrm{cal} = \mathrm{SX}_\mathrm{raw} + \Delta$. The corrections are linear (and therefore continuous) functions of the following form:
\begin{eqnarray}
\Delta \teff & = & a_0 + a_1 \left(\frac{\teff - 5500~\K}{100~\K}\right), \nonumber\\
\Delta \logg & = & b_0 + b_1 \left(\frac{\logg - 3.5~\dex}{0.1~\dex}\right), \nonumber\\
\Delta \feh  & = & c_0 + c_1 \left(\frac{\feh}{0.1~\dex}\right), \nonumber
\end{eqnarray}
where $a_0 = -28.7~\K$ and $a_1  = -5.17$; $b_0 = 0.0146~\dex$, $b_1 = 0.0028$, $c_0 = 0.0034~\dex$, and $c_1 = -0.0175$.   After applying these corrections, we compare the calibrated \SME and \SpecMatch parameters in Figure~\ref{fig:sm-sx}. We find a dispersion of \SXTeffRMS~K, \SXLoggRMS~dex, and \SXFehRMS~dex in \teff, \logg, \feh, respectively.

\subsection{Parameter Averaging}
\label{sec:averaging}

\begin{figure*}
\centering
\includegraphics[width=0.8\textwidth]{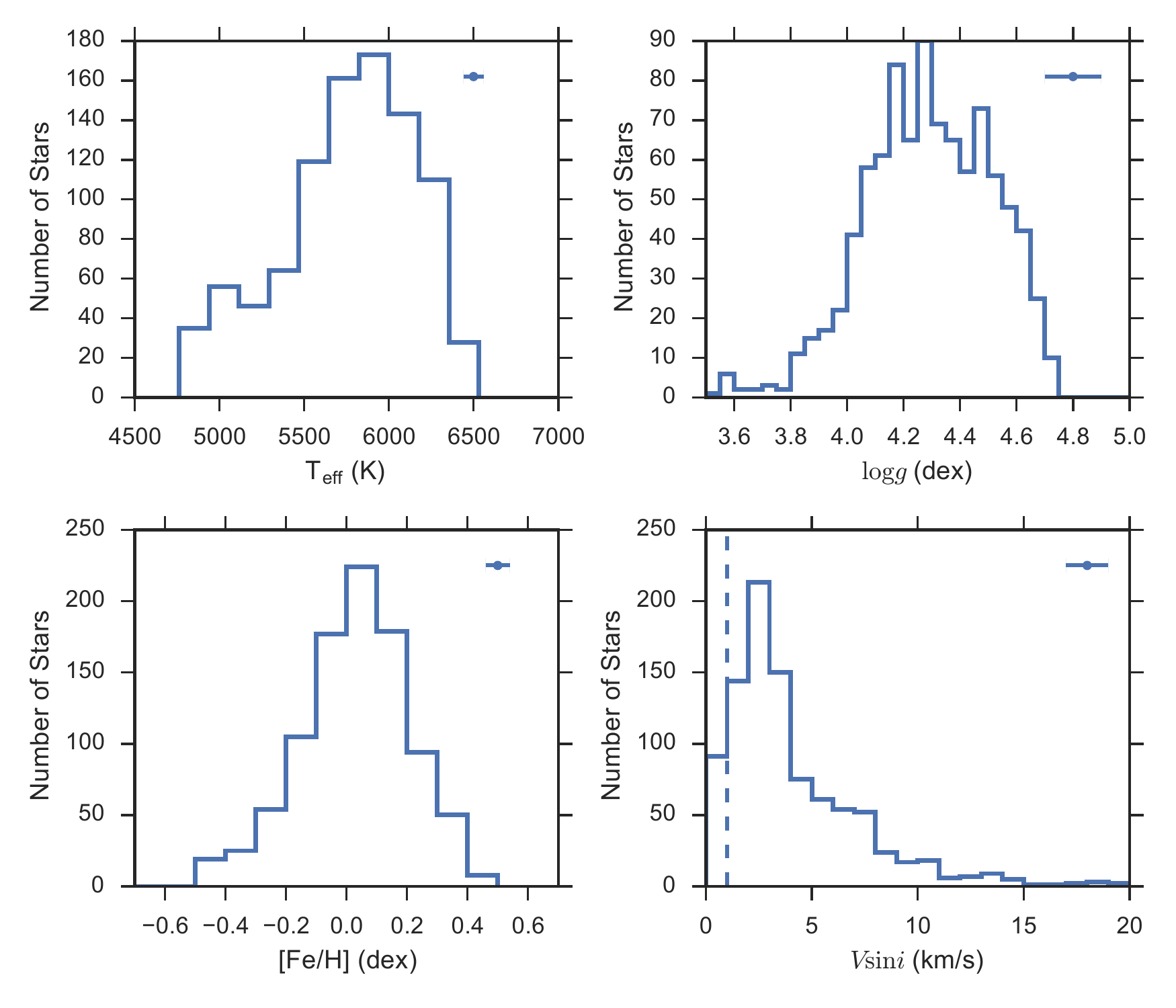}
\caption{Histograms of the adopted spectroscopic parameters (\teff, \logg, \feh and \vsini) for all stars in our CKS sample.  Adopted uncertainties (Table \ref{tab:uncertainties}) are plotted in the upper right corner of each panel. \vsini\ is difficult to measure for the most slowly rotating stars.  Thus we adopt 2 \kms as an upper limit for stars with reported \vsini $<$ 1 \kms\ (dashed line).\label{fig:four_panel_props}}
\end{figure*}

One of the key features of the CKS catalog is that three quarters of the spectra (\nstarsSX/\nstars) were analyzed by two independent spectral analysis pipelines. This enables the straightforward identification of suspect spectroscopic parameters where the two techniques produce disparate results.  For stars with consisent parameters from both catalogs, we adopted the arithmetic mean the \SpecMatch and \SME values for \teff, \logg, and \feh.  We adopted the \SpecMatch \vsini\ values for all stars.  For a small number of stars, we rejected the parameter values determined by one of the two pipelines; these cases are described below. The distributions of adopted values of \teff, \logg, \feh, and \vsini are shown in Figure \ref{fig:four_panel_props}.

\subsection{Outlier Rejection}
\label{sec:outliers}
\begin{figure*}
\centering
\includegraphics[width=0.8\textwidth]{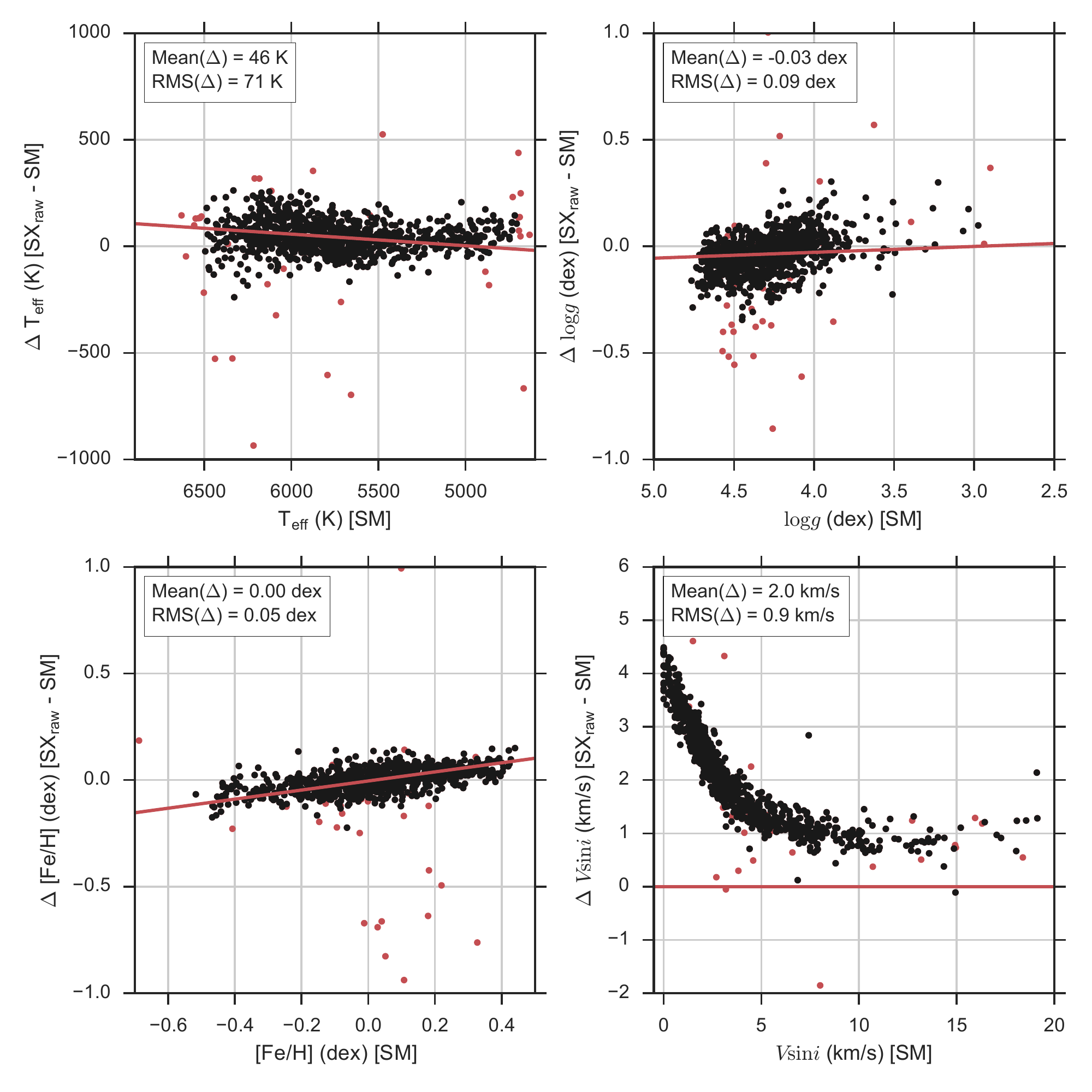}
\caption{Four panels showing the differences in stellar parameters determined independently by the \SpecMatch (SM) and \SME ($\mathrm{SX}_\mathrm{raw}$) algorithms. Panels correspond to \teff (upper left), \logg (upper right), \feh (lower right), and \vsini (lower right).  Each panel shows the difference between the SM and raw SX parameter values for each star, as a function of the SM values.  Annotations give the mean and RMS differences between the SM and uncalibrated SX catalogs. Red lines show the corrections that were applied to SX parameter values (see Sec.\ \ref{sec:calibration-sme}). Subsequent figures show SX parameter values with these corrections applied. We have highlighted the \nstarsDisagree stars where significant disagreement exists between the two methods see Sec.~\ref{sec:outliers}. These stars are excluded from the calibrations and subsequent analyses.\label{fig:sx-on-sm}}
\end{figure*}

Determining parameters from two pipelines offers the opportunity to discover cases of significant disagreement. We identify \nstarsDisagree stars where any of the following conditions are satisfied: (1) \teff differed by more than 300~K, (2) \logg differed by more than 0.35~dex, or (3) \feh differed by more than 0.30~dex. These stars are highlighted in Figure~\ref{fig:sx-on-sm} and are marked with a flag in the machine-readable version of Table~\ref{tab:cks}. We recommend excluding these stars from further statistical analyses. 

\begin{figure*}
\centering
\includegraphics[width=0.8\textwidth]{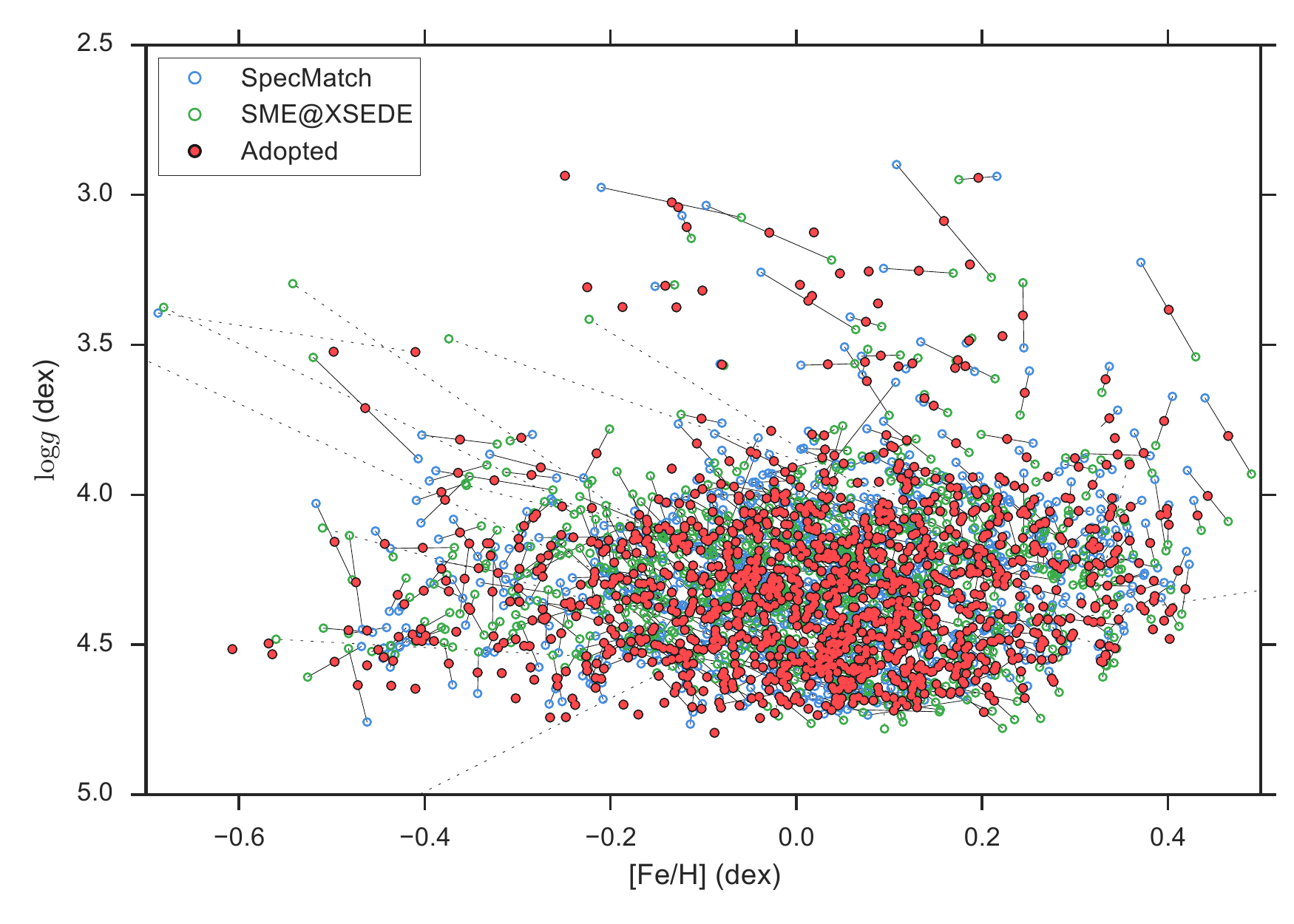}
\caption{Same as Figure\ \ref{fig:cks-dumbbell-steff-smet} except the axes are \feh\ and \logg.\label{fig:cks-dumbbell-smet-slogg}}
\end{figure*}

In anticipation of studies where a preferred value for one or more of these stars is needed, we inspected the parameters from \SpecMatch and \SME. In cases where one pipeline clearly failed, we adopted the triplet of parameters (\teff, \logg, \feh) determined by the other method. Figures \ref{fig:cks-dumbbell-steff-smet}--\ref{fig:cks-dumbbell-smet-slogg} show the spectroscopic parameters in three planes.  Outliers can be identified by dashed lines with red points marking the adopted values.  

The KIC \citep{Brown2011_kic} offers a third determination of effective temperature (see Sec.\ \ref{sec:compare-cks-kic}).  For cases of $>$300 K disagreement between \SpecMatch and \SME, we adopted \teff from the pipeline closest matching to the KIC value. We adopted the \SpecMatch parameters for KOI-156, KOI-719, KOI-935, KOI-3683, KOI-4060. For KOI-870, we choose the \SpecMatch value because the mean stellar density determined from the transit light curve is nearer to that implied by the \SpecMatch parameters. For KOI-1054, we adopted the \SME value because those parameters more closely match the KIC parameters.

For cases where \logg disagreed by >0.35~dex, we adopted the parameter that most closely matched a previously published result, when available. We adopted \SpecMatch values for KOI-3, KOI-104, KOI-1963, KOI-4601, KOI-4651, KOI-4699.

For cases with significant \logg disagreement, but lacking existing literature values, we inspected the combinations of \teff and \logg returned by \SpecMatch and \SME and searched for cases where one pipeline gave values that are inconsistent with the observed properties of normal stars (e.g. \citealt{Torres2010}). We adopt \SpecMatch parameters for the following stars KOI-2287, KOI-2503, and KOI-3928 because the \SME parameters constitute unphysical combinations of \teff and \logg.

Finally, for KOI-193, KOI-2228, KOI-2481, KOI-2676, KOI-2786, KOI-3203, KOI-3215, KOI-3419, and KOI-4053 there was no clear indication of a failure in either of the pipelines, so we simply averaged the parameters.

\subsection{Adopted Values}
\label{sec:adpoted}

Table \ref{tab:cks} lists the adopted values \teff, \logg, \feh, and \vsini, 
as well as individual determinations by the \SpecMatch and \SME pipelines.  
We also list radial velocities relative to the barycenter of the solar system, having accuracies of 0.1 \kms, determined 
using the method of \cite{Chubak2012}. 

\subsection{Precise Validation with Platinum Sample}
All methods to determine spectroscopic parameters have some systematic and random errors.  
We use two methods, asteroseismology and line-by-line spectroscopic synthesis, as validation standards against which we calibrate the CKS results. These results are summarized in Table \ref{tab:comp_other_surveys}.

\subsubsection{Huber et al.\ (2013)}
\label{sec:huber13}
\begin{figure*}
\includegraphics[width=1\linewidth]{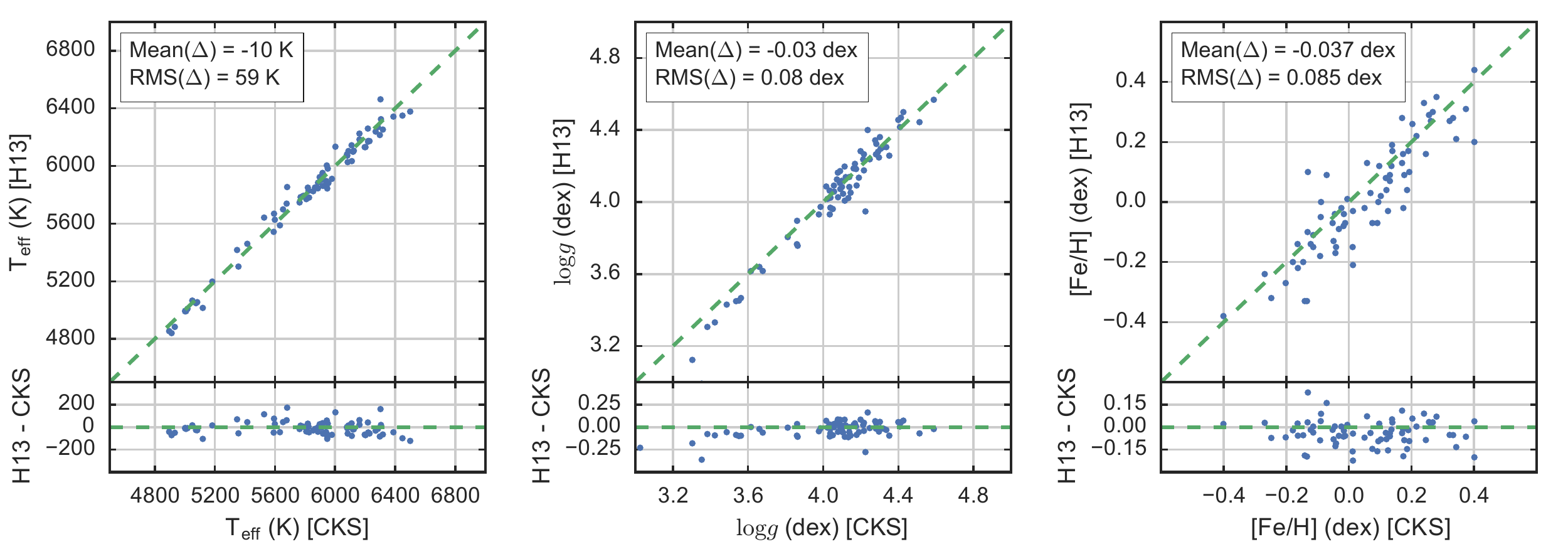}
\caption{Comparison of \teff (left), \logg\ (middle), and \feh\ (right) values between CKS and \cite{Huber_2013} (H13) asteroseismic analysis for \HuberBNstarscommon\ stars in common. Annotations indicate the mean and RMS differences between the samples.\label{fig:compare-cks-huber13}}
\end{figure*}

\cite{Huber_2013} measured the properties of 77 planet host stars using \Kepler asteroseismology. The asteroseismic analysis is much more precise than our spectroscopic method in \logg determination and is only modestly sensitive to the input values of \teff and \feh, which were measured by the SPC method \citep{Buchhave2012}.  As described in \cite{Petigura2015}, we used \HuberBNstarscommon of the stars in the \cite{Huber_2013} sample to compare with our CKS results. Figure \ref{fig:compare-cks-huber13} compares the spectroscopic parameters for the stars in common between CKS and \cite{Huber_2013}. We find excellent agreement in \logg with an offset of \HuberBLoggOffset dex and an RMS of \HuberBLoggRMS dex between the two measurement techniques.  This tight agreement between asteroseismology and CKS supports the \cksSiglogg~dex adopted uncertainty for the CKS \logg values.

For the lowest gravity stars in the comparison, we note a systematic trend in $\Delta \logg$. At \logg = 3.2~dex, the CKS gravities are 0.2~dex larger than the \cite{Huber_2013} values. This trend may be due in part to discrepancies between the B16 spectroscopic gravities and asteroseismic gravities for evolved stars. B16 demonstrated 0.05~dex (RMS) agreement with asteroseismology for a sample of 42 \Kepler stars with \logg = 3.7--4.5~dex. Thus, the B16 gravities may be offset from asterosiesmic gravities for stars with \logg < 3.7~dex. This systematic trend affects only small subset of the CKS sample. The vast majority (97\%) the stars are high gravity ($\logg > 3.7$~dex), where we see excellent agreement with asteroseismology. 


\subsubsection{Bruntt et al. (2012)}
\label{sec:brunnt12}

As a second validation sample, we used the results for the 93 ``platinum stars'' identified and analyzed by the \Kepler Project to establish stellar parameters of the highest possible accuracy. These 93 stars are all bright and were subjects of asteroseismic and spectroscopic analyses. \cite{Bruntt2012} (B12) gathered high-resolution ($R$ = 80,000), high S/N (200-300 per pixel) spectra of these solar type stars using the ESPaDOnS spectrograph on the 3.6-m Canada-France-Hawaii Telescope.  They used the VWA \citep{Bruntt2010} analysis tool to perform an iterative, line-by-line spectroscopic synthesis to match the observed spectra.  This tool has itself been calibrated on samples with asteroseismic and interferometric measurements.  The spectroscopic fits were done with \logg\ held fixed to values determined by asteroseismic analysis of \Kepler photometry \citep{Verner2011b,Verner2011a}.  

\begin{figure*}
\includegraphics[width=1\linewidth]{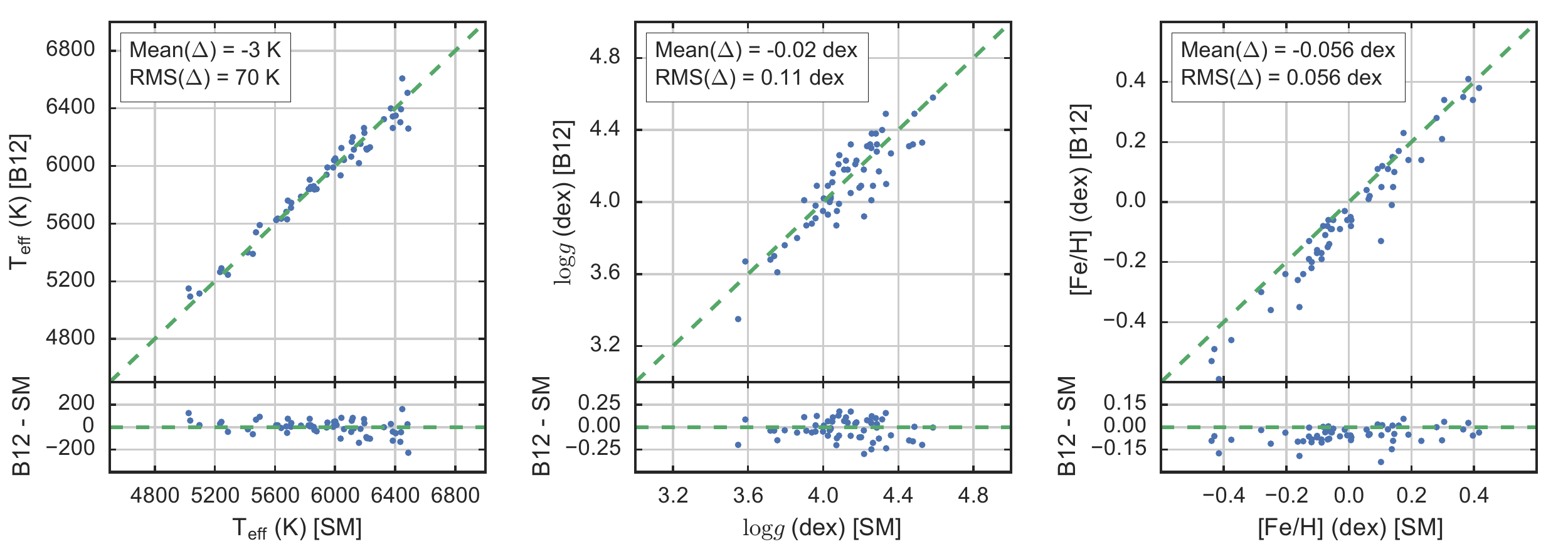}
\caption{Comparison of \teff (left), \logg\ (middle), and \feh\ (right) values between \SpecMatch (SM) and \cite{Bruntt2012} (B12) for \BrunttSMNstarscommon\ stars in common.  Annotations indicate the mean and RMS differences between the samples.\label{fig:compare-sm-bruntt12}}
\end{figure*}

Figure~\ref{fig:compare-sm-bruntt12} compares the spectroscopic parameters for \BrunttSMNstarscommon\ stars in common between \SpecMatch and (B12).  Note that these stars are generally not the hosts of transiting planets, and thus are not part of the CKS sample.  The HIRES spectra for this comparison were gathered separately.  The parameters \teff, \logg\, and \feh\ all show good agreement with negligible offsets and low scatter.  This establishes the precision and accuracy of \SpecMatch and CKS (see Sec.\ \ref{sec:uncertainties} and Table \ref{tab:uncertainties}).

\subsection{Uncertainties}
\label{sec:uncertainties}

We adopt a precision of 60 K for \teff for comparison within this catalog. This is based on the 60~K agreement between \SpecMatch and \cite{Brewer16} (B16) temperatures.  Because of systematic differences between \teff scales between catalogs (see e.g. \citealt{Pinsonneault2012,Brewer16}), we encourage adding 100~K systematic uncertainty in quadrature (116 K total uncertainty) for applications beyond internal comparisons within the CKS catalog. 

\begin{figure*}
\centering
\includegraphics[width=1.0\textwidth]{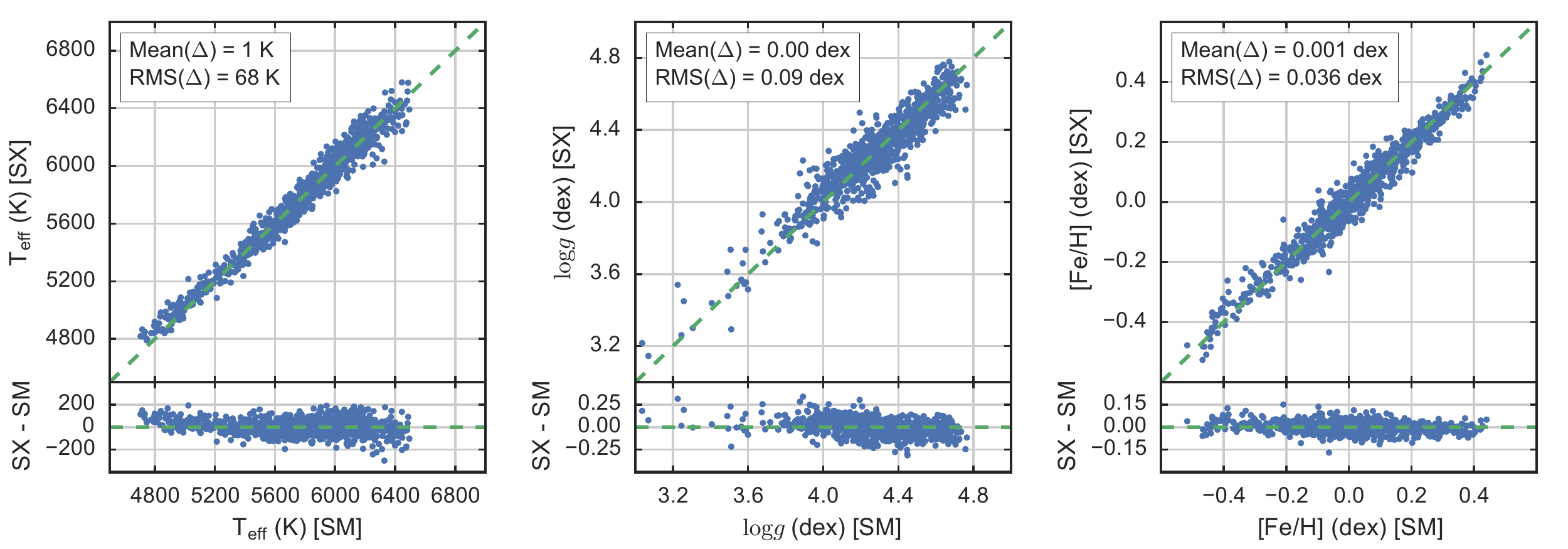}
\caption{Comparison of \SpecMatch (SM) and \SME (SX) values for \teff, \logg, and \feh. The \SME values have been adjusted to the \SpecMatch scale (Sec.\ \ref{sec:averaging}). The top panel compares SM and SX parameters while the lower panel shows their difference as a function of the SM parameters. Equality between SM and SX are shown as green lines. The RMS value is the standard deviation of difference between SM and SX values for the same star.\label{fig:sm-sx}}
\end{figure*}

We adopt a \logg\ uncertainty in this catalog of 0.10 dex based on the agreement between \SpecMatch and B16 surface gravities. This is supported by the \SXLoggRMS~dex agreement between \SpecMatch and \SME gravities (Figure~\ref{fig:sm-sx}) as well as the agreement with asteroseismic gravities, presented in Sections~\ref{sec:huber13} and \ref{sec:brunnt12}.

For spectroscopic analyses, modeling uncertainties such as incomplete or inaccurate line lists, imperfect model atmospheres, and the assumption of LTE will influence the derived \teff, \logg, and \feh.  For \teff and \logg, there are independent measurement techniques that yield parameters with precisions and accuracies that are comparable to, or higher than, those from spectroscopy. Examples include the Infrared Flux Method (IRFM) for \teff and asteroseismology for \logg. These independent techniques are often used to characterize the modeling uncertainties associated with spectroscopy.
 
Characterizing the effect of modeling uncertainties on spectroscopic metallicities is challenging because there are no non-spectroscopy techniques with comparable precision/accuracy that can serve to validate the spectroscopic metallicities. A standard method to quantify such errors is to compare metallicities derived through different codes with the assumption that the model-dependent uncertainties are reflected in the scatter and offsets between the two techniques.
 
We note the agreement between metallicities derived through four different techniques that all analyzed high resolution, high SNR spectra. \SpecMatch, \SME, B16, and B12  used a variety of line lists, radiative transfer codes, and model atmospheres. We observe a 0.036 dex scatter between \SpecMatch and \SME metallicities and a 0.06 dex scatter between \SpecMatch and B16 metallicities.
 
The metallicities of both \SpecMatch and \SME were placed onto the B16 scale, so there are no mean offsets by construction. However, in comparing SM to B12, we note a slight deviation from the 1-to-1 line and a mean offset of 0.056 dex. This reflects different metallicity scales associated with the B16 and B12 analyses, which likely stem from different line lists, model atmospheres, radiative transfer codes, etc. 
 
We adopt a metallicity precision of 0.04 dex for comparison within this catalog motivated by the \SpecMatch-\SME agreement.  Because of systematic differences between the B16 and B12 metallicity scales, we encourage adding 0.06~dex systematic uncertainty in quadrature (0.07 dex total uncertainty) for applications beyond internal comparisons within the CKS catalog. 

The \vsini values are entirely determined from \SpecMatch.  We adopt 1-$\sigma$ errors of 1 \kms\ and an upper limit of 2 \kms\ for stars with \vsini\ $<$ 1 \kms.  This uncertainty is based on a comparison of \vsini values determined by Rossiter-McLaughlin measurements  \citep{Albrecht2012} to the \SpecMatch-determined values for the same stars \citep{Petigura2015}.


\begin{figure*}
\centering
\includegraphics[width=0.8\textwidth]{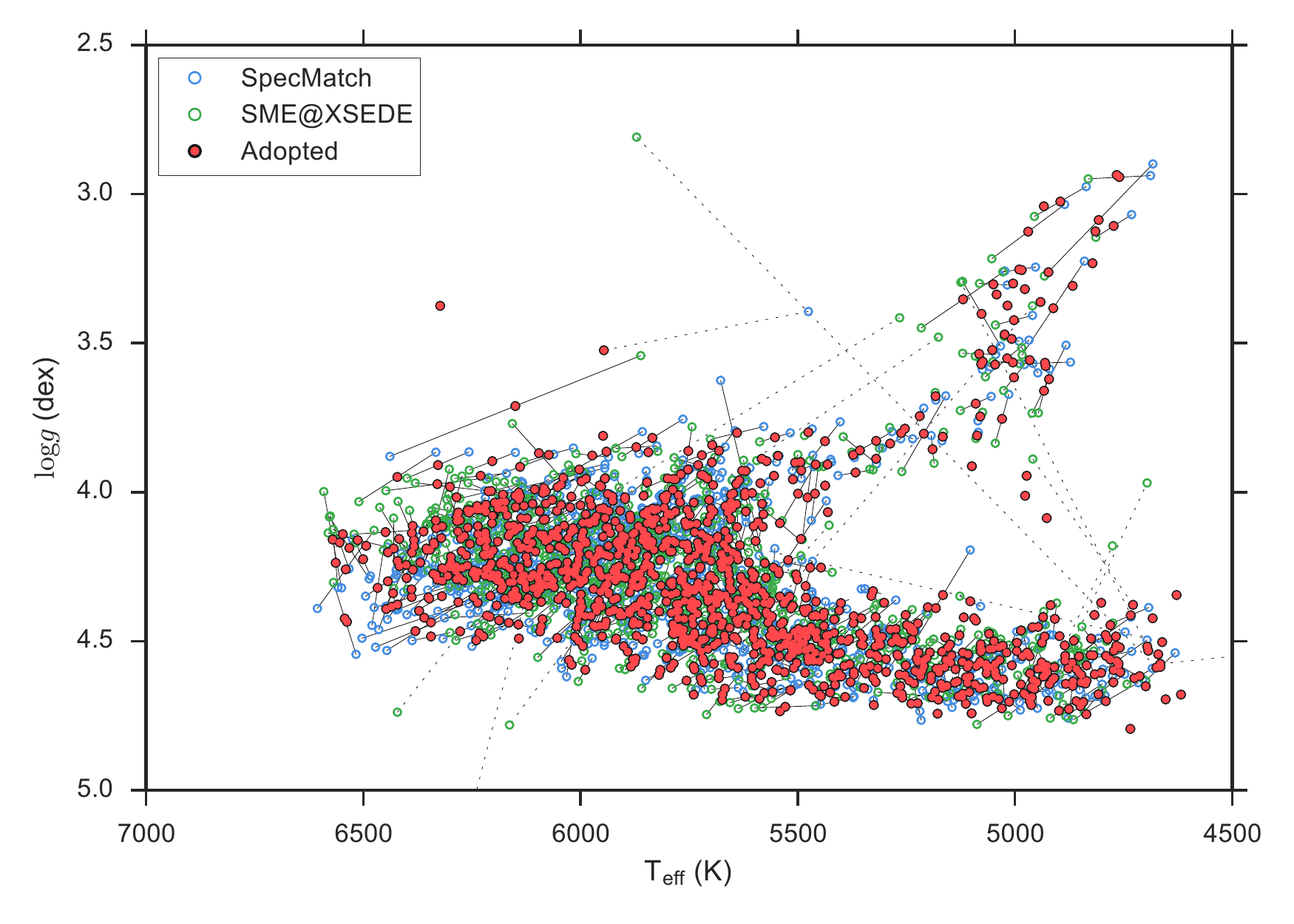}
\caption{Hertzsprung-Russell diagram (\logg\ versus \teff) for CKS stars. Blue points are parameter values from \SpecMatch, green are from \SME, and red are the adopted values. Solid lines connect \SpecMatch and \SME values for the same star, for cases in which simple averaging of the results of the two methods was applied. Dashed lines connect \SpecMatch and \SME values for which the results of one method was rejected and the other was adopted. \SME values have been corrected to be on the \SpecMatch scale (Sec.~\ref{sec:calibration-sme}).\label{fig:cks-dumbbell-steff-smet} }
\end{figure*}

\begin{figure*}
\centering
\includegraphics[width=0.8\textwidth]{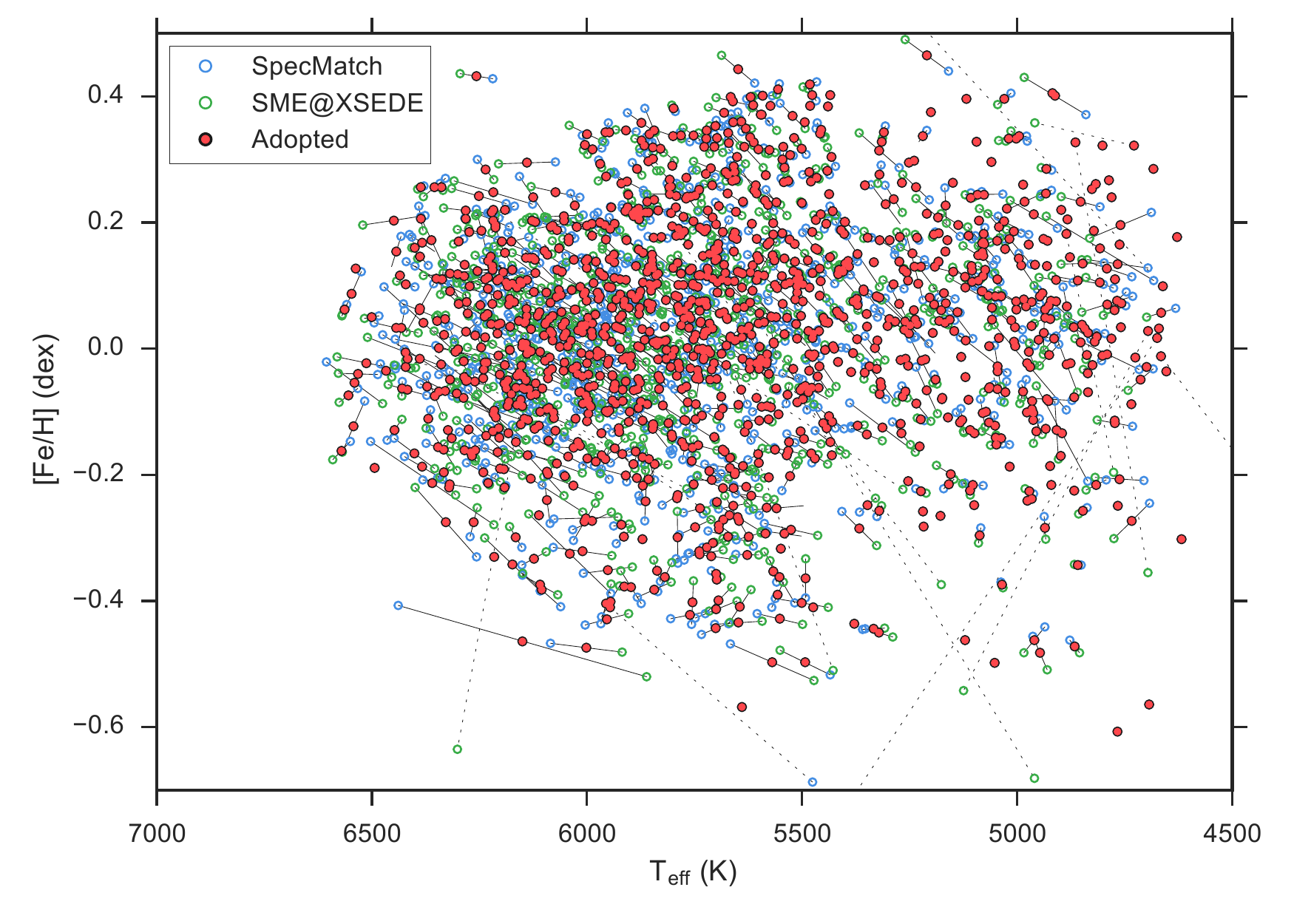}
\caption{Same as Figure\ \ref{fig:cks-dumbbell-steff-smet} except the axes are \teff and \feh.\label{fig:cks-dumbbell-steff-smet}}
\end{figure*}

\begin{deluxetable*}{lrrrrrrrrrrrrrrr}[tb]
\tabletypesize{\scriptsize}
\tablecaption{Spectroscopic Parameters}
\tablewidth{0pt}
\tablehead{
   \colhead{} & \multicolumn{4}{c}{Adopted Values}  & \colhead{} & \multicolumn{4}{c}{SpecMatch}  & \colhead{} & \multicolumn{4}{c}{\SME} \\ 
 \cline{2-5}  \cline{7-10}   \cline{12-15} \\
  \colhead{KOI}   & 
  \colhead{\teff} & 
  \colhead{\logg}  & 
  \colhead{\feh} & 
  \colhead{\vsini}  &
  \colhead{}   & 
  \colhead{\teff} & 
  \colhead{\logg}  & 
  \colhead{\feh} & 
  \colhead{\vsini}  &
  \colhead{}   & 
  \colhead{\teff} & 
  \colhead{\logg}  & 
  \colhead{\feh} & 
  \colhead{\vsini} & 
  \colhead{TRV} \\
  \colhead{No.} & 
  \colhead{(K)} & 
  \colhead{(dex)}  & 
  \colhead{(dex)} & 
  \colhead{ (\kms)} &
  \colhead{}   & 
  \colhead{(K)} & 
  \colhead{(dex)}  & 
  \colhead{(dex)} & 
  \colhead{ (\kms)} &
  \colhead{}   & 
  \colhead{(K)} & 
  \colhead{(dex)}  & 
  \colhead{(dex)} & 
  \colhead{(\kms)} &
  \colhead{(\kms)}
}
K00001 & 5819 & 4.40 & $+$0.01 & 1.3 &  & 5853 & 4.43 & $+$0.02 & 1.3 &  & 5785 & 4.37 & $+$0.01 & 4.3 & $+$0.5\\ 
K00002 & 6449 & 4.13 & $+$0.20 & 5.2 &  & 6376 & 4.13 & $+$0.21 & 5.2 &  & 6521 & 4.14 & $+$0.20 & 6.1 & $-$10.4\\ 
K00003 & 4864 & 4.50 & $+$0.33 & 3.2 &  & 4864 & 4.50 & $+$0.33 & 3.2 &  & 4696 & 3.97 & $-$0.36 & 3.1 & $-$63.4\\ 
K00006 & 6348 & 4.36 & $+$0.04 & 11.8 &  & 6348 & 4.36 & $+$0.04 & 11.8 &  & \nodata & \nodata & \nodata & \nodata & $-$42.8\\ 
K00007 & 5827 & 4.09 & $+$0.18 & 2.8 &  & 5813 & 4.03 & $+$0.17 & 2.8 &  & 5841 & 4.15 & $+$0.18 & 4.6 & $-$60.8\\ 

\enddata
\tablecomments{Adopted Values are our best determination of the spectroscopic parameters after calibrating the \SME values and averaging with the \SpecMatch values.  Uncertainties for the Adopted Values are summarized in Table~\ref{tab:uncertainties} and Section~\ref{sec:uncertainties}. Results from \SME (after the calibrations described in Section~\ref{sec:calibration-sme}) and \SpecMatch are also presented. This table will be published in its entirety in the machine-readable format in the accepted version of this paper. A portion is shown here for guidance regarding its form and content.\label{tab:cks}}
\end{deluxetable*}

\begin{deluxetable}{ll}
\tabletypesize{\footnotesize}
\tablecaption{Adopted Parameter Uncertainties
\label{tab:uncertainties}}
\tablewidth{0pt}
\tablehead{
  \colhead{Parameter}   & 
  \colhead{1-$\sigma$ Uncertainty}
}
\startdata
\teff & $\pm$60 K (relative; within this catalog) \\
      & $\pm$100 K (systematic) \\
\logg & $\pm$0.10 dex \\
\feh  & $\pm$0.04 dex (relative; within this catalog) \\
      & $\pm$0.04 dex (systematic) \\
\vsini & $\pm$1 \kms  \\
       & $<$ 2 \kms upper limit for \vsini\ $<$ 1 \kms 
\enddata
\end{deluxetable}

\section{Comparison with Other Surveys of Kepler Planet Hosts}
\label{sec:comparison}
Table \ref{tab:comp_other_surveys} provides a comparison between CKS results and several surveys of KOIs, described below.

\subsection{Kepler Input Catalog}
\label{sec:compare-cks-kic}

The Kepler Input Catalog \citep[KIC;][]{Brown2011_kic} was constructed prior to the launch of \Kepler from $griz$ $+$ Mg b photometry.  It was well suited for the purpose of selecting appropriate stars to be monitored by the spacecraft photometer.  The KIC has stated uncertainties of 200 K in \teff and 0.4 dex in \logg (both for \teff in the range 4500-6500 K).  Metallicity ($\log(Z)$) was reported, but the uncertainties were expected to be high.%
\footnote{
\cite{Brown2011_kic} states, ``it is difficult to assess the reliability of our log($Z$) estimates, but there is reason to suspect that it is poor, particularly at extreme \teff.''
}
While the KIC was used with great success to select dwarf Sun-like stars for the mission, it did not provide reliable surface gravity and metallicity measurements.  This was one of the primary motivations of the CKS project.
Figure \ref{fig:compare-sm-bruntt12} compares the CKS stellar parameters to the KIC.

\subsection{Huber et al. (2014)}
\label{sec:compare-cks-huber14}

\begin{figure*}
\centering
\includegraphics[width=1\linewidth]{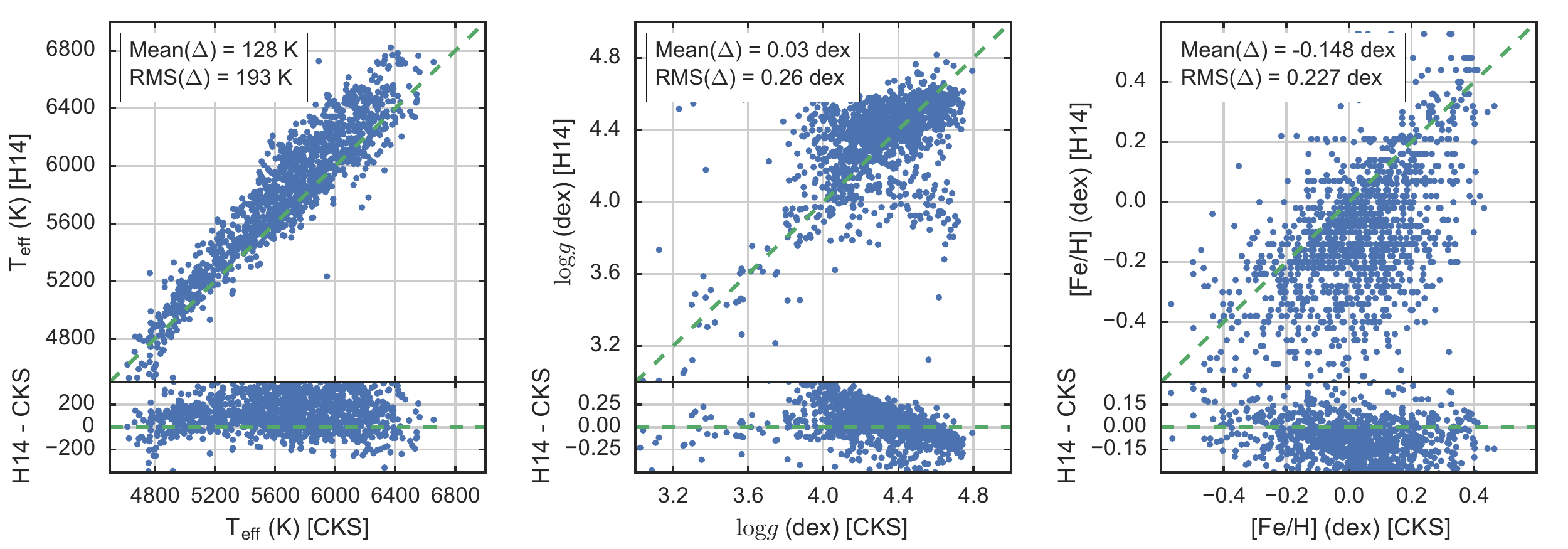}
\caption{Comparison of \teff (left), \logg\ (middle), and \feh\ (right) values between CKS and the revised stellar properties from the \Kepler team \citep[H14;][]{Huber_2014} for \HuberNstarscommon\ stars in common.  Annotations indicate the mean and RMS differences between the samples.}
\label{fig:compare-cks-huber14}
\end{figure*}

\cite{Huber_2014} provided a comprehensive update to the KIC by compiling literature measurements of stellar properties from different observational techniques (photometry, spectroscopy, asteroseismology, and exoplanet transits) and homogeneously fitting them to a grid of Dartmouth stellar isochrones.  This often allowed the uncertainties in the stellar parameters to be reduced, in comparison to the KIC.   For the 1244 stars analyzed by \cite{Huber_2014} for which we have spectroscopy, their stated uncertainties are 2--3.5\% (fractional) in \teff, 0.40 dex to 0.15 dex in \logg, and 0.30 to 0.15 dex in \feh, all considerably larger than the CKS errors here.
Figure \ref{fig:compare-cks-huber14} compares the \SpecMatch and  \cite{Huber_2014} values.

\subsection{LAMOST}
\label{sec:compare-cks-lamost}

\begin{figure*}
\epsscale{1.0}
\includegraphics[width=1\linewidth]{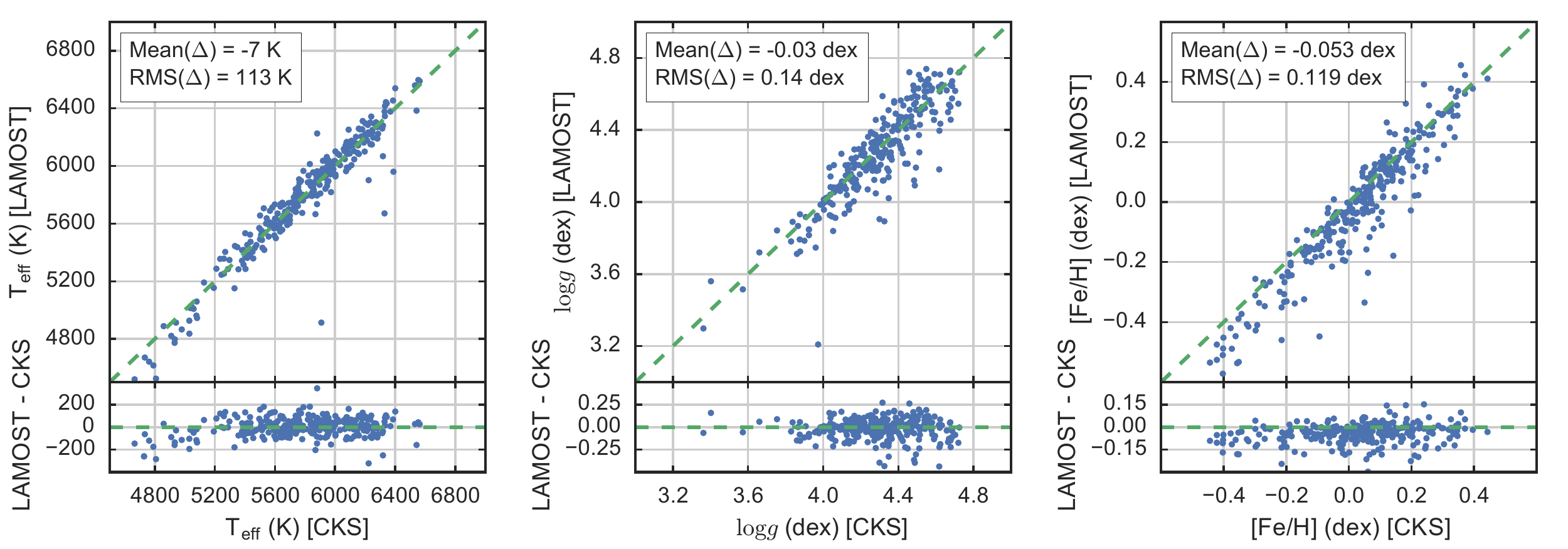}
\caption{Comparison of \teff (left), \logg\ (middle), and \feh\ (right) values between CKS and the LAMOST survey \citep{deCat2015} for \LAMOSTNstarscommon\ stars in common.  Annotations indicate the mean and RMS differences between the samples.}
\label{fig:compare-cks-lamost}
\end{figure*}

The Large Sky Area Multi-Object Fiber Spectroscopic Telescope \citep[LAMOST;][]{Luo2012,Dong2014} is instrumented with highly-multiplexed (4000 fibers per 5 degree field), low-resolution ($R$ = 1000 or 5000) spectrometer.  It can cover the entire \Kepler Field in 14 pointings.  
LAMOST is engaged in several large spectroscopic surveys, including a set of 6500 asteroseismic targets and $\sim$150,000 ``planet targets'' in the \Kepler Field.  
The LAMOST Stellar Parameter \citep[LASP;][]{Luo2012,Wu2014} pipeline is used to compute \teff, \logg,\ and \feh, which are stored in a large catalog \citep{deCat2015}.  
The stated uncertainties for LAMOST are typically 100 K in \teff, 0.10 dex in \logg, and 0.10 dex in \feh.
In Figure \ref{fig:compare-cks-lamost} we compare \SpecMatch and LAMOST results.

\subsection{SPC}
\label{sec:comp_spc}

\begin{figure*}
\includegraphics[width=1\linewidth]{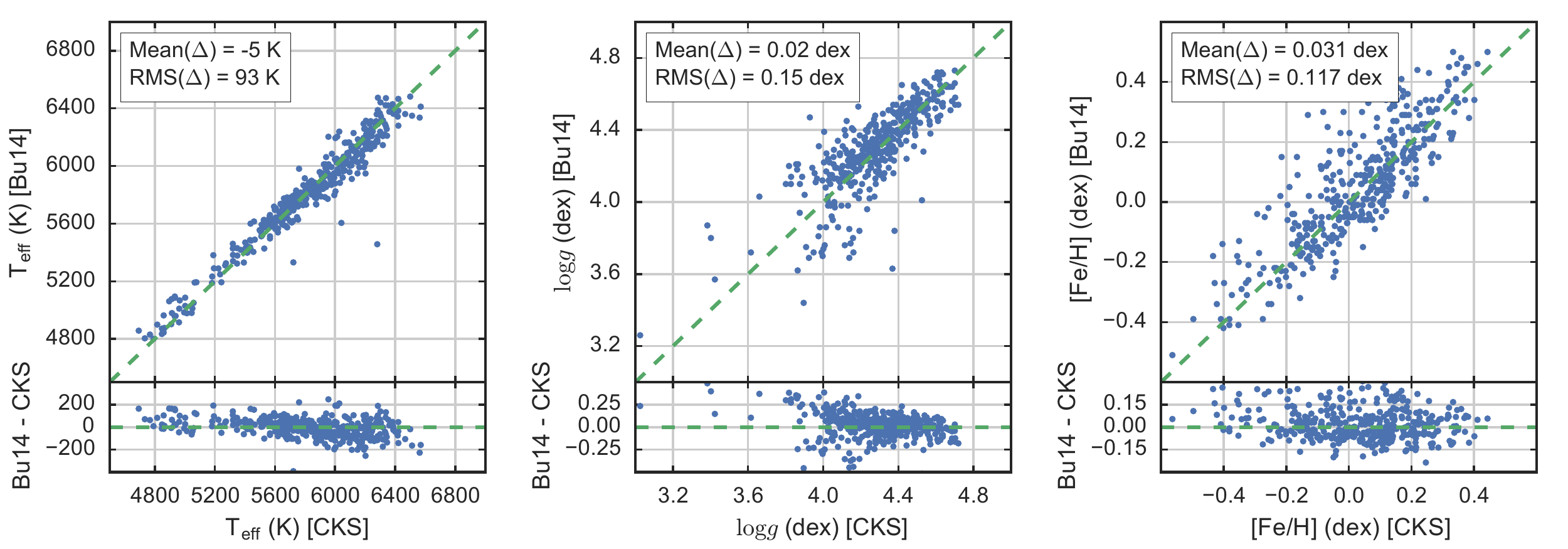}
\caption{Comparison of \teff (left), \logg\ (middle), and \feh\ (right) values between CKS and analysis for high-resolution spectroscopy using SPC \citep[Bu14;][]{Buchhave_2014} for \BuchhaveNstarscommon\ stars in common.  Annotations indicate the mean and RMS differences between the samples.}
\label{fig:compare-cks-buch14}
\end{figure*}

The Stellar Parameter Classification \citep[SPC;][]{Buchhave2012,Buchhave_2014,Buchhave2015} tool matches observed high-resolution spectra to a library grid of synthetic model spectra using a prior on \logg from stellar evolutionary models.  The stated uncertainties for SPC are typically 50 K in \teff, 0.10 dex in \logg, and 0.08 dex in \feh.  Figure \ref{fig:compare-cks-buch14} compares \SpecMatch and SPC results.  The SPC results are from FIES, TRES, and HIRES spectra \citep{Buchhave_2014}.

\subsection{KEA}
\label{sec:comp_kea}

\begin{figure*}
\epsscale{1.0}
\includegraphics[width=1\linewidth]{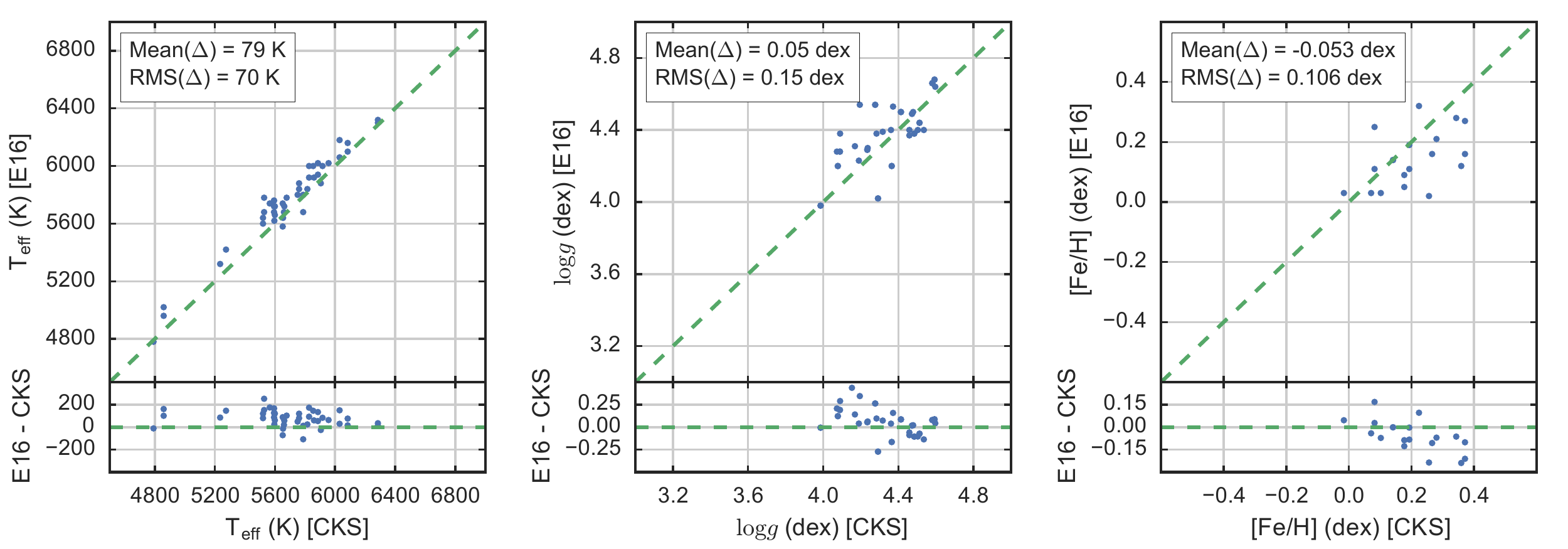}
\caption{Comparison of \teff (left), \logg\ (middle), and \feh\ (right) values between CKS and analysis for high-resolution spectroscopy using KEA \citep[E16;][]{Endl2016} for \EndlNstarscommon\ stars in common.  Annotations indicate the mean and RMS differences between the samples.}
\label{fig:compare-cks-endl16}
\end{figure*}

KEA \citep{Endl2016} is a spectral analysis tool that uses a large grid of model stellar spectra \citep{Kurucz1993} computed with an LTE spectrum synthesis  \citep{Sneden1973}.
KEA was calibrated on \Kepler ``platinum stars'' and has stated uncertainties of 200 K in \teff, 0.18 dex in \logg, and 0.12 dex in \feh.  Figure \ref{fig:compare-cks-endl16} compares results from \SpecMatch and KEA-analyzed spectra from McDonald Observatory.  The comparison with CKS is limited in usefulness because of only \EndlNstarscommon\ stars in common that span a relatively narrow range of \logg\ and \feh.

\subsection{Everett et al. (2013)}
\label{sec:compare-cks-everett13}

\begin{figure*}
\epsscale{1.0}
\includegraphics[width=1\linewidth]{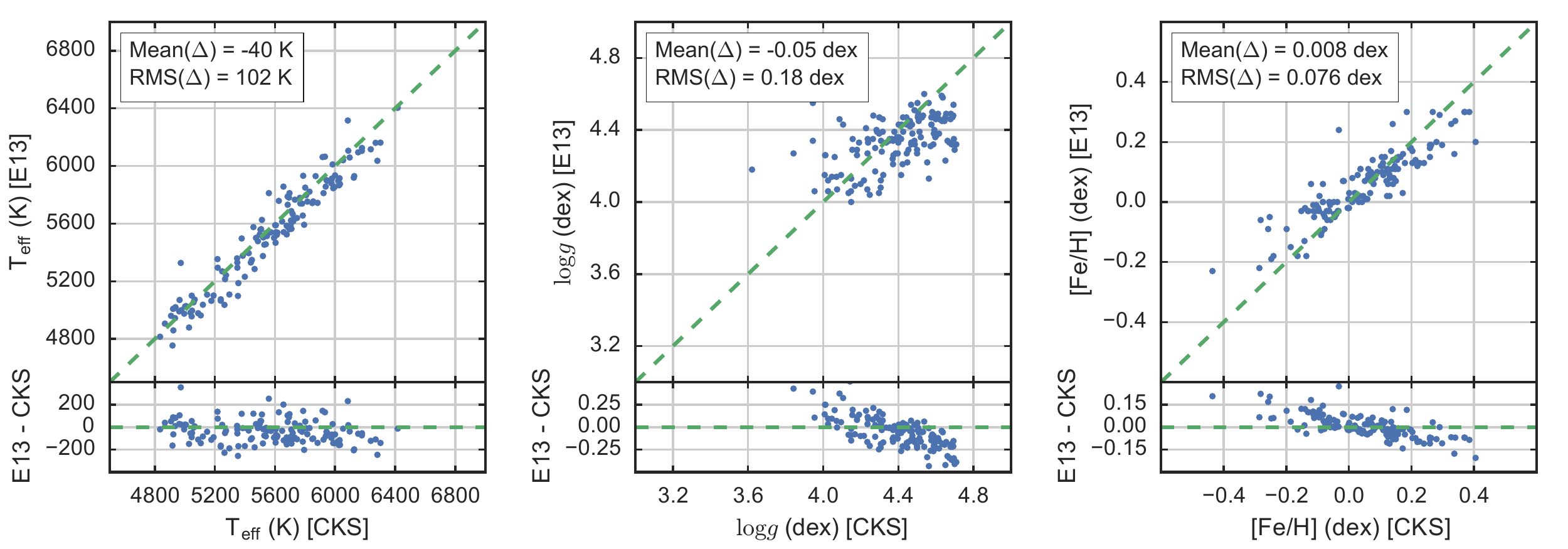}
\caption{Comparison of \teff (left), \logg\ (middle), and \feh\ (right) values between CKS and analysis for medium-resolution spectroscopy by \cite[E13;][]{Everett2013} for \EverettNstarscommon\ stars in common.  Annotations indicate the mean and RMS differences between the samples.}
\label{fig:compare-cks-everett13}
\end{figure*}

\cite{Everett2013} measured low-resolution ($R = 3000$) optical spectra of 268 stars using the National Optical Astronomy Observatory (NOAO) Mayall 4 m telescope on Kitt Peak and the facility RCSpec long-slit spectrograph.  They report uncertainties of 75 K in \teff, 0.15 dex in \logg, and 0.10 dex in \feh.  Figure \ref{fig:compare-cks-everett13} compares CKS and \cite{Everett2013}.  Note the systematic trends in \logg and \feh in the comparison plots.

\subsection{Flicker}
\label{sec:compare-cks-bastien14-logg}

\begin{figure}
\centering
\includegraphics[width=0.8\linewidth]{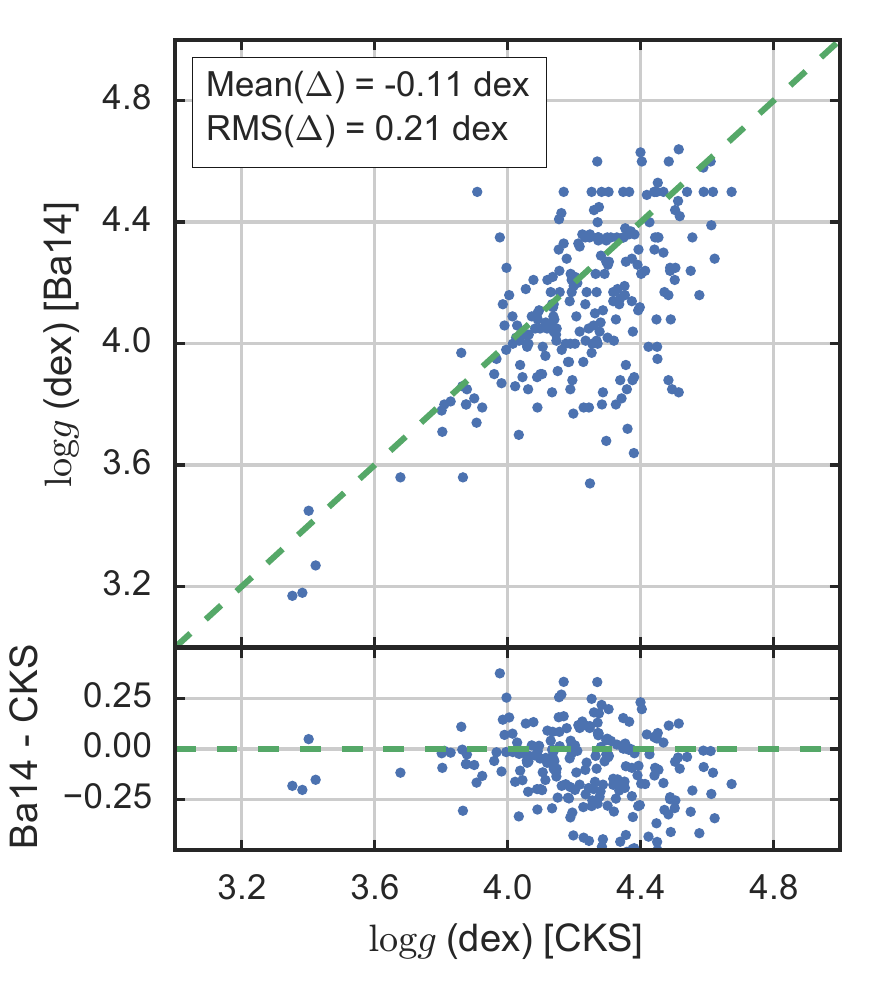}
\caption{Comparison of \logg\ values between CKS and the Flicker method \cite[Ba14;][]{Bastien2014a} for \BastienNstarscommon\ stars in common.  Annotations indicate the mean and RMS differences between the samples.}
\label{fig:compare-cks-bastien14-logg}
\end{figure}

\cite{Bastien2013} developed a method to measure \logg using \Kepler light curves themselves. ``Flicker'' measures photometric variability from convective granulation on short timescales. It works because the amplitude of convective granulation depends on the strength of the restoring force, i.e., surface gravity.  \cite{Bastien2014a} noted that Flicker-based gravities were systematically higher than those in the KIC, implying that most \Kepler planets (which lacked spectroscopically-determined gravities) had radii that were underestimated by 20-30\%. \cite{Bastien2016} improved the Flicker method by measuring photometric variability on multiple timescales, but excluded KOIs from their catalog.  Figure \ref{fig:compare-cks-bastien14-logg} compares CKS and \cite{Bastien2014a} \logg performance for stars brighter than \Kp = 13.  As noted in \cite{Bastien2016}, Flicker performs best for the brightest stars with the lowest photon-limited noise.


\begin{figure*}
\centering
\includegraphics[width=1\linewidth]{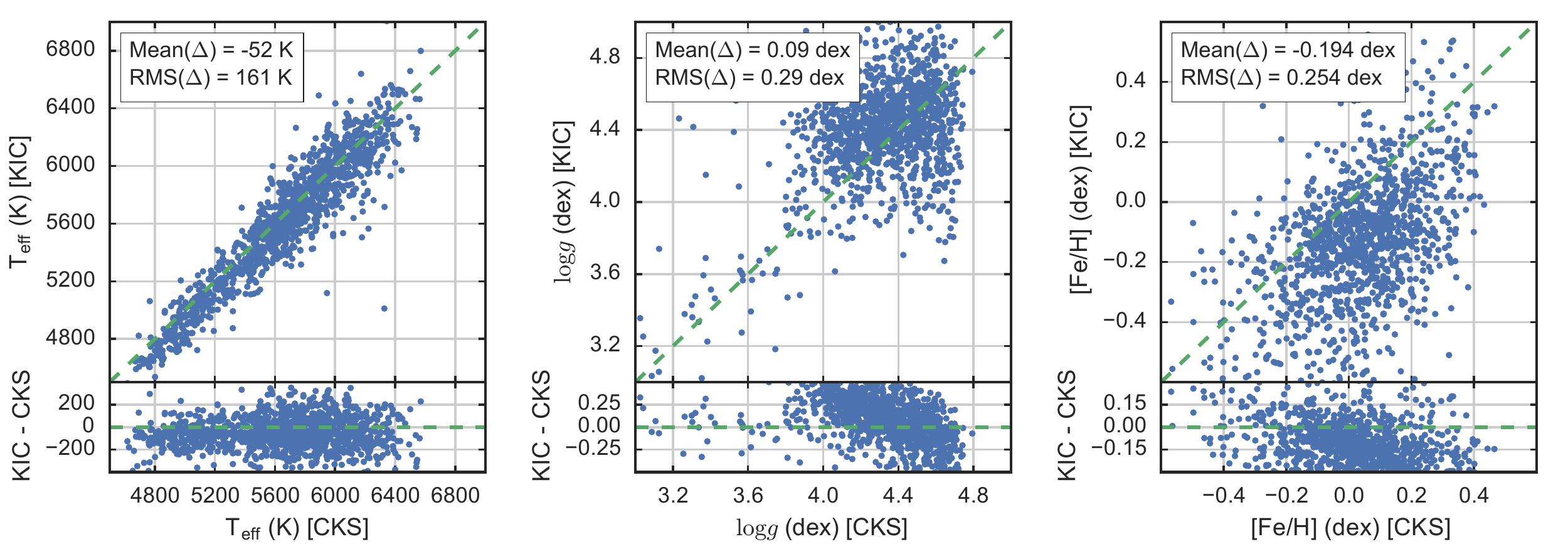}
\caption{Comparison of \teff (left), \logg\ (middle), and \feh\ (right) values between CKS and Kepler Input Catalog \citep[KIC;][]{Brown2011_kic} for \KICNstarscommon\ stars in common.  Annotations indicate the mean and RMS differences between the samples.}
\label{fig:compare-cks-kic}
\end{figure*}

\begin{deluxetable*}{lccccccccccc}
\tabletypesize{\footnotesize}
\tablecaption{Comparison with Other Surveys
\label{tab:comp_other_surveys}}
\tablewidth{0pt}
\tablehead{
  \colhead{} & \multicolumn{3}{c}{Stated Uncertainties} & \colhead{} & \multicolumn{3}{c}{Offset with CKS} & \colhead{} & \multicolumn{3}{c}{RMS with CKS} \\ 
  \cline{2-4} \cline{6-8} \cline{10-12}  \\
  \colhead{Catalog}   & 
  \colhead{\teff} &
  \colhead{\logg}  &
  \colhead{\feh}  &
  \colhead{$N_\mathrm{\star}$}  &
  \colhead{\teff} &
  \colhead{\logg}  &
  \colhead{\feh} &
  \colhead{} &
  \colhead{\teff} &
  \colhead{\logg}  &
  \colhead{\feh} \\
  \colhead{}   & 
  \colhead{[K]} &
  \colhead{[dex]}  &
  \colhead{[dex]}  &
  \colhead{common}  &
  \colhead{[K]} &
  \colhead{[dex]}  &
  \colhead{[dex]}  &
  \colhead{} &
  \colhead{[K]} &
  \colhead{[dex]}  &
  \colhead{[dex]}  
}
\startdata
\sidehead{This Paper}
~~~~~~CKS \tablenotemark{a} & \specialcell{\cksSigTeff~(rel)\\100  (sys)}  & \cksSiglogg & \specialcell{\cksSigfeh~(rel)\\0.06  (sys)}  & \nodata & \nodata & \nodata & \nodata & & \nodata & \nodata & \nodata\\
\sidehead{Validation of CKS with Platinum Stars}
~~~~~~\cite{Huber_2013} \tablenotemark{b} & \nodata  & 0.01  & \nodata & \HuberBNstarscommon & \nodata & \HuberBLoggOffset & \nodata & & \nodata & \HuberBLoggRMS & \nodata\\
~~~~~~\cite{Bruntt2012} \tablenotemark{c} & 60  & 0.03  & 0.06 & \BrunttSMNstarscommon & \BrunttSMTeffOffset & \BrunttSMLoggOffset & \BrunttSMFehOffset & & \BrunttSMTeffRMS & \BrunttSMLoggRMS & \BrunttSMFehRMS\\
\sidehead{Comparison Surveys}
~~~~~~KIC \citep{Brown2011_kic} \tablenotemark{d} & 200  & 0.40 & $\sim$0.30  & \KICNstarscommon & \KICTeffOffset & \KICLoggOffset & \KICFehOffset & & \KICTeffRMS & \KICLoggRMS & \KICFehRMS\\
~~~~~~\cite{Huber_2014} \tablenotemark{e} & \specialcell{110 (sp)\\193 (ph)}  & \specialcell{0.15 (sp)\\0.40 (ph)}  & \specialcell{0.15 (sp)\\0.30 (ph)} & \HuberNstarscommon & \HuberTeffOffset & \HuberLoggOffset & \HuberFehOffset & & \HuberTeffRMS & \HuberLoggRMS & \HuberFehRMS\\
~~~~~~LAMOST \citep{deCat2015} & 100 & 0.10 & 0.10 & \LAMOSTNstarscommon & \LAMOSTTeffOffset & \LAMOSTLoggOffset & \LAMOSTFehOffset & & \LAMOSTTeffRMS & \LAMOSTLoggRMS & \LAMOSTFehRMS\\
~~~~~~SPC \citep{Buchhave_2014} & 50  & 0.10  & 0.08 & \BuchhaveNstarscommon & \BuchhaveTeffOffset & \BuchhaveLoggOffset & \BuchhaveFehOffset & & \BuchhaveTeffRMS & \BuchhaveLoggRMS & \BuchhaveFehRMS\\
~~~~~~KEA \citep{Endl2016} & 100  & 0.18  & 0.12  & \EndlNstarscommon & \EndlTeffOffset & \EndlLoggOffset & \EndlFehOffset & & \EndlTeffRMS & \EndlLoggRMS & \EndlFehRMS\\
~~~~~~\cite{Everett2013} & 75  & 0.15  & 0.10 & \EverettNstarscommon & \EverettTeffOffset & \EverettLoggOffset & \EverettFehOffset & & \EverettTeffRMS & \EverettLoggRMS & \EverettFehRMS\\
~~~~~~Flicker \citep{Bastien2014a} \tablenotemark{f} & \nodata  & 0.10  & \nodata & \BastienNstarscommon & \nodata & \BastienLoggOffset & \nodata & & \nodata & \BastienLoggRMS & \nodata
\enddata
\tablenotetext{a}{CKS uncertainties in \teff are 60 K within the sample (rel) and 100 K systematic uncertainty (sys) when compared to other surveys.}
\tablenotetext{b}{\cite{Huber_2013} is a platinum sample for \logg\ measurements only, using asteroseismology.  \teff\ and \feh\ for this sample are based on SPC; see Sec.\ \ref{sec:huber13}.}
\tablenotetext{c}{The comparison with \cite{Bruntt2012} is with \SpecMatch parameters, not \SME or their combination, CKS.}
\tablenotetext{d}{Errors for the KIC are for \teff in the range 4500--6500 K. }
\tablenotetext{e}{Errors for \cite{Huber_2014} are specified separately for stars with photometry (ph) or also spectroscopy (sp).  \teff errors are stated as 3.5\% (193 K at 5500 K) for photometry and 2\% (110 K at 5500 K) for spectroscopy.}
\tablenotetext{f}{Flicker \logg uncertainties are higher that 0.10 dex for some stars.}
\end{deluxetable*}

\section{Summary and Discussion}
\label{sec:summary}

We present precise stellar parameters (\teff, \logg, \feh, and \vsini) for \nstars \Kepler planet host stars based on a uniform set of high-S/N, high-resolution spectra from Keck/HIRES. Our magnitude-limited (\Kp $<$ 14.2) CKS sample, augmented with multi-planet systems and other planet samples, constitutes the largest set of stars and transiting planets with precisely determined stellar parameters to date. 

Stellar parameters were determined using two methods, \SpecMatch and \SME.  The zero-points and scales of our measurements are calibrated against ``platinum star'' samples observed with higher precision methods (asteroseismology and line-by-line spectral synthesis applied to high-S/N spectra).  The uncertainties of our adopted parameters are \cksSigTeff~K in \teff, \cksSiglogg~dex in \logg, \cksSigTeff~dex in \feh, and 1 \kms in \vsini.  

We find that the \Kepler planet host stars have distributions of \teff, \logg, \feh, and \vsini and an H-R Diagram that are similar to those of stars in the solar neighborhood, given the selection effects from the planet detection process of \Kepler.  In particular, for the magnitude-limited sample (\Kp $<$ 14.2), our CKS parameters give a median metallicity for  \Kepler planet host stars of $-0.01$ dex and an RMS of 0.19 dex.  
\cite{Valenti05} measured the solar neighborhood to have a median metallicity of 0.00 dex and an RMS of 0.24 dex.

Additional CKS papers build on this work.  
In Paper II (Johnson et al., submitted), we compute precise stellar radii and masses, and approximate stellar ages. 
Paper III (Fulton et al., submitted) examines the planet radius distribution using our precise stellar radii and the planet-to-star radius ratios from \Kepler photometry.  
Paper IV (Petigura et al., to be submitted) examines the metallicities of stars in the CKS sample.  
Paper V (Weiss et al., to be submitted) probes the similarities and differences in planetary and stellar properties for single and multi-planet transiting systems.

{\it Facilities:} \facility{Keck:I (HIRES)}, \facility{Kepler}

\acknowledgments{
The CKS project was conceived, planned, and initiated by AWH, GWM, JAJ, HTI, and TDM.  
AWH, GWM, JAJ acquired Keck telescope time to conduct the magnitude-limited survey.  
Keck time for the other stellar samples was acquired by JNW, LAR, and GWM.
The observations were coordinated by HTI and AWH and carried out by AWH, HTI, GWM, JAJ, TDM, BJF, LMW, EAP, ES, and LAH. AWH secured CKS project funding. \SpecMatch was developed and run by EAP and \SME was developed and run by LH and PAC. Downstream data products were developed by EAP, HTI, and BJF. Results from the two pipelines were consolidated and the integrity of the parameters were verified by AWH, HTI, EAP, GWM, with assistance from BJF, LMW, ES, LAH, and IJMC. This manuscript was largely written by AWH and EAP with significant assistance from HTI, JNW, and GWM.

We thank Jason Rowe, Dan Huber, Jeff Valenti, Natalie Batalha, and David Ciardi for helpful conversations 
and Roberto Sanchis-Ojeda for his work on the Ultra-Short Period planet sample.
We thank the many observers who contributed to the measurements reported here. 
PAC and LH thank Jeff Valenti, and Eric Stempels for their extensive help in running SME and developing the SME implementation presented in this paper. 
We gratefully acknowledge the efforts and dedication of the Keck Observatory staff, especially Randy Campbell, Scott Dahm, Greg Doppmann, Marc Kassis, Jim Lyke, Hien Tran, Josh Walawender, Greg Wirth for support of HIRES and of remote observing. Most of the data presented herein were obtained at the W.\ M.\ Keck Observatory, which is operated as a scientific partnership among the California Institute of Technology, the University of California, and NASA. We are grateful to the time assignment committees of the University of Hawaii, the University of California, the California Institute of Technology, and NASA for their generous allocations of observing time that enabled this large project.
Kepler was competitively selected as the tenth NASA Discovery mission. Funding for this mission is provided by the NASA Science Mission Directorate.  
We thank the Kepler Science Office, the Science Operations Center, Threshold Crossing Event Review Team (TCERT), and the Follow-up Observations Program (FOP) Working Group for their work on all steps in the planet discovery process ranging from selecting target stars and pointing the \Kepler telescope to developing and running the photometric pipeline to curating and refining the catalogs of \Kepler planets.  We specifically thank Natalie Batalha, William Borucki, and David Ciardi in particular, for selecting stars in the Habitable Zone sample.
EAP acknowledges support from Hubble Fellowship grant HST-HF2-51365.001-A awarded by the Space Telescope Science Institute, which is operated by the Association of Universities for Research in Astronomy, Inc.\ for NASA under contract NAS 5-26555. 
AWH acknowledges NASA grant NNX12AJ23G.  
TDM acknowledges NASA grant NNX14AE11G.
PAC acknowledges National Science Foundation grant AST-1109612.
LH acknowledges National Science Foundation grant AST-1009810.
LMW acknowledges support from Gloria and Ken Levy and from the Trottier Family.
ES is supported by a post-graduate scholarship from the Natural Sciences and Engineering Research Council of Canada.
IJMC performed his work under contract with the Jet Propulsion Laboratory (JPL) funded by NASA through the Sagan Fellowship Program executed by the NASA Exoplanet Science Institute.
Work by JNW was partly supported by a NASA Keck PI Data Award, administered by the NASA Exoplanet Science Institute.
This work made use of the SIMBAD database (operated at CDS, Strasbourg, France), NASA's Astrophysics Data System Bibliographic Services, and the NASA Exoplanet Archive, which is operated by the California Institute of Technology, under contract with the National Aeronautics and Space Administration under the Exoplanet Exploration Program.
LAR gratefully acknowledges support provided by NASA through Hubble Fellowship grant \#HF-51313 awarded by the Space Telescope Science Institute, which is operated by the Association of Universities for Research in Astronomy, Inc., for NASA, under contract NAS 5-26555. This work was performed in part under contract with the Jet Propulsion Laboratory (JPL) funded by NASA through the Sagan Fellowship Program executed by the NASA Exoplanet Science Institute.
Finally, the authors wish to recognize and acknowledge the very significant cultural role and reverence that the summit of Maunakea has always had within the indigenous Hawaiian community.  We are most fortunate to have the opportunity to conduct observations from this mountain.}

\bibliographystyle{apj}
\bibliography{manuscript}

\end{document}